\documentclass[twocolumn,aps,pra,floatfix,superscriptaddress]{revtex4-2}
\usepackage[braket, qm]{qcircuit}
\usepackage{mathptmx}
\usepackage{amsmath}
\usepackage{amssymb}
\usepackage{braket}
\usepackage{graphicx}
\usepackage{amsfonts}
\usepackage{csquotes}
\usepackage{hhline}
\usepackage{relsize}
\usepackage{listings}
\usepackage{color}
\usepackage{qcircuit}
\usepackage{physics}
\usepackage{soul}
\usepackage[most]{tcolorbox}
\usepackage{booktabs}

\usepackage{adjustbox}

\usepackage{float}
\usepackage{tikz}
\usetikzlibrary{positioning}
\lstdefinestyle{mystyle}{
    backgroundcolor=\color{backcolour},   
    commentstyle=\color{codegreen},
    keywordstyle=\color{magenta},
    numberstyle=\tiny\color{codegray},
    stringstyle=\color{codepurple},
    basicstyle=\footnotesize,
    breakatwhitespace=false,         
    breaklines=true,                 
    captionpos=b,                    
    keepspaces=true,                 
    numbers=left,                    
    numbersep=5pt,                  
    showspaces=false,                
    showstringspaces=false,
    showtabs=false,                  
    tabsize=2
}
 
\usepackage{subfigure}

\usepackage{dcolumn}
\usepackage{tabularx}
\usepackage[colorlinks=true,linkcolor=blue,citecolor=blue,urlcolor=blue]{hyperref}

\usepackage{dcolumn}
\usepackage{cases}
\usepackage{multirow}
\usepackage{xcolor}
\makeatletter
\let\newfloat\newfloat@ltx
\makeatother

\usepackage{algorithm,algorithmic}

\begin{document}

\title{A Resource Efficient Quantum Kernel} 

\author{Utkarsh Singh}
\affiliation{National Research Council of Canada, 100 Sussex Drive, Ottawa, Ontario K1N 5A2, Canada}
\affiliation{Department of Physics, University of Ottawa, 25 Templeton Street, Ottawa, Ontario, K1N 6N5 Canada}

\author{Jean-Fr\'ed\'eric Laprade}
\affiliation{Institut quantique, Université de Sherbrooke Sherbrooke, QC J1K 2R1, Canada}
\author{Aaron Z. Goldberg}
\affiliation{National Research Council of Canada, 100 Sussex Drive, Ottawa, Ontario K1N 5A2, Canada}
\author{Khabat Heshami}
\affiliation{National Research Council of Canada, 100 Sussex Drive, Ottawa, Ontario K1N 5A2, Canada}
\affiliation{Department of Physics, University of Ottawa, 25 Templeton Street, Ottawa, Ontario, K1N 6N5 Canada}
 \affiliation{Institute for Quantum Science and Technology, Department of Physics and Astronomy, University of Calgary, Alberta T2N 1N4, Canada}

\begin{abstract}

Quantum processors may enhance machine learning by mapping high-dimensional data onto quantum systems for processing. Conventional feature maps for encoding data onto a quantum circuit are impractical, as the number of entangling gates scales quadratically with the dimension of the dataset. We introduce a quantum feature map designed to handle high-dimensional data with a significantly reduced number of qubits and entangling operations. Our approach preserves essential data characteristics while promoting computational efficiency, evidenced by extensive experiments on benchmark datasets that demonstrate a marked improvement in both accuracy and resource utilization when using our feature map as a kernel for characterization, as compared to state-of-the-art quantum feature maps. Our noisy simulation results highlight our map's ability to function within the constraints of noisy intermediate-scale quantum devices. Through numerical simulations and small-scale implementation on a superconducting circuit quantum computing platform, we demonstrate that our scheme performs on par or better than a set of classical algorithms for classification. Our approach empirically delays the onset of exponential concentration relative to existing feature maps under the diagnostics considered here, which can improve the practical operating regime of fidelity kernels on near-term devices. Our results indicate that resource-efficient feature maps can broaden the range of kernel-based QML experiments that are feasible on near-term platforms, and motivate further validation on larger datasets and additional hardware.
\end{abstract}

\maketitle

\section{Introduction \label{intro}}

High-dimensional data is prevalent in modern machine learning tasks, including image and speech recognition, natural language processing, and medical diagnostics. While classical machine learning techniques can handle these high-dimensional problems, they often require substantial computational resources, particularly as the dimensionality of the data increases~\cite{Chen2009, Berisha2021Oct, Bellman1961, Bishop2006PRML, VerleysenFrancois2005Curse}. Quantum computing has emerged as a promising avenue to address these challenges. Quantum kernel methods, for instance, have shown potential for accelerating data analysis by efficiently learning relationships in high-dimensional spaces encoded into quantum states~\cite{Havlicek2019Mar, Peters2021Nov, Kusumoto2021Jun, Wu2021Sep, Alaminos2022Feb, Lloyd2014Sep}. To harness large-dimensional Hilbert spaces for processing data, quantum computers need to efficiently encode classical data onto quantum states. These quantum feature maps are required for quantum support vector classification \cite{Biamonte2017Sep, Schuld2015Apr, Schuld2019Feb, Schuld2021Jan, qiskit-textbook} and quantum neural networks \cite{Biamonte2017Sep, Schuld2015Apr, cong2019quantum, zoufal2019quantum,qiskit-textbook}, including data reuploading \cite{PerezSalinas2020datareuploading}, and are germane to essentially all quantum machine learning paradigms. Yet, the most popular feature maps require numbers of qubits and controlled-not (CNOT) gates linear and quadratic in the number of data features, respectively, severely shortening their practical application. We introduce a feature map whose qubit requirement is reduced by a factor of at least two and whose CNOT-gate requirement is reduced to linear in the number of qubits, and show these improvements to be sufficient for practical application on current devices.

 Quantum feature maps~\cite{Schuld2019Feb} translate classical data into a quantum form that can be manipulated using quantum circuits, enabling the application of quantum algorithms~\cite{Schuld2015Apr}. 
The qubit and entangling gate requirements for conventional feature maps pose a significant limitation for practical applications, particularly in era of noisy intermediate-scale quantum (NISQ) devices, where the number of qubits is limited and gate errors are prohibitively large~\cite{Preskill2018Aug}. The development of efficient quantum feature maps that can effectively encode high-dimensional classical data onto quantum states while minimizing the number of qubits remains an active area of research, with several proposed embedding schemes under exploration~\cite{Havlicek2019Mar, Lloyd2014Sep, Hubregtsen2021Mar}.

Here, after introducing our quantum feature map with the reduced resource requirements, we examine its performance using a kernel method in classification of various datasets. We compare our results with several classical methods and the commonly used Pauli feature map known as the ZZ feature map and consistently show competing performance.We emphasize that ZZFeatureMap is used here as a widely available reference implementation of a Pauli-style entangling feature map, rather than as the only relevant baseline. \cite{Havlicek2019Mar, ZZAbbas2021Jun, lungcancer2025qsvm, alzheimer2023qml, benchmarking2024kernel, Boy_2025, bonddissociation2025adapting, anastomoticleak2025qml, anomaly2024additive, imbalance2025leveraging, comparative2023study, genomic2025modeling} Many alternative embeddings exist (including data re-uploading and other hardware-efficient constructions), and to avoid over-interpreting the comparison as ``CPMap vs.\ ZZ'', we additionally benchmark CPMap against a suite of fixed (non-trained) data re-uploading feature maps in Appendix~\ref{data_reuploading_fm}.
We present a detailed background of quantum feature maps and their challenges in Section~\ref{secII}. In Section~\ref{secIII} we introduce our approach. We lay out our experimental methodology and results in Section~\ref{secIV}, including small-scale tests on IBM's ibm\_quebec and IBM\_torino machines. Our results stimulate considerations of quantum machine learning as an early practical application of application-starved NISQ devices. In our work, we utilize this feature map within a predefined quantum kernel function known as a the fidelity quantum kernel for the support vector classifier (SVC) algorithm in Qiskit \cite{Qiskit, qiskit-textbook}; as a result, we will refer to both the feature map and the kernel as CPMap.

\subsection{Feature Maps \label{secII}}

Quantum feature maps (QFMs) play a pivotal role in quantum machine learning, enabling the encoding of classical data into the quantum state space. This encoding process transforms classical vectors into the amplitudes of quantum states, thus facilitating quantum processing. In this section, we delve into the mathematical foundations of QFMs and explore their applications and limitations.

\subsubsection{Mathematical Definition}
A quantum feature map is defined as a function $\Phi: \mathbb{R}^n \rightarrow \mathcal{H}$, where $\mathcal{H}$ is the Hilbert space of a quantum system. Given a classical vector $\mathbf{x} \in \mathbb{R}^n$, the map $\Phi$ transforms $\mathbf{x}$ into a quantum state:

\begin{equation}
\Phi(\mathbf{x})=U_\Phi(\mathbf{x})|0\rangle = \sum_{i=0}^{2^n - 1} f_i(\mathbf{x}) |i\rangle,
\end{equation}
where $f_i(\mathbf{x})$ are QFM-dependent functions of the classical data and $\{|i\rangle\}$ represents the orthonormal set of computational basis states. The most common class of feature maps comprises the Pauli feature maps. 

\subsubsection{Pauli Feature Maps}

The Pauli feature map for encoding classical data into the state space of a quantum system is formalized as
\begin{equation}
U_{PFM}(\mathbf{x}) = \exp \left( i \sum_{S \subseteq T} \phi_S(\mathbf{x}) \prod_{i \in S} P_i \right)H^{\otimes n}.
\end{equation}
Here, \( S \) indexes subsets of qubits, \( T \) encompasses all such subsets, and each \( P_i \) represents one of the Pauli matrices \(\{I, X, Y, Z\}\) acting on the qubit labeled by $i$. The factor $H^{\otimes n}$ applies a Hadamard gate $H$ on each of the $n$ qubits to transform computational basis states into superpositions thereof. The function \( \phi_S \) is defined as the data-mapping function, with \( \phi_S(\mathbf{x}) = x_i \) when \( S \) is a singleton and \( \phi_S(\mathbf{x}) = \prod_{j \in S} (\pi - x_j) \) otherwise, thus capturing both individual feature impacts and the impacts of higher-order interactions between features within the quantum state \cite{Havlicek2019Mar, Vasques2023Jul}; in this sense the scalar functions $\phi(\mathbf{x})$ are components of the mapping function $\Phi$. 

Among the commonly used Pauli feature maps is the \textit{Pauli-Z Feature Map} (also referred to as the ZFeatureMap), where each subset \( S \) is a singleton and each \( P_i \) corresponds to a Pauli-Z operation. This map encodes classical data by applying phase shifts relative to each data feature \( x_i \), such that each qubit in the circuit experiences a phase rotation as dictated by the respective feature value. The corresponding transformation is expressed as
\begin{equation}
U_{Z}(\mathbf{x}) = \bigotimes_{i} e^{-i x_i Z} H,
\end{equation}
where the tensor product runs over all qubits. The result is a quantum state separable between all qubits where for each qubit a data feature is encoded in the relative phase between computational basis states \cite{Havlicek2019Mar, qiskit-textbook}.

A second widely adopted feature map is the \textit{Pauli-ZZ Feature Map} (ZZFeatureMap), which extends the Pauli-Z map by incorporating interactions between qubits. This is achieved through alternating single-qubit Z-rotations and two-qubit ZZ-entangling gates. The latter take the form
\begin{equation}
U_{ZZ}(\mathbf{x}) = \exp \left( i \sum_{\{ i, j \} \in T} \phi_{\{i,j\}}(\mathbf{x}) Z_i Z_j \right) H^{\otimes n}
\end{equation}
and enable the map to encode both individual contributions and pairwise interactions between features into an entangled quantum state. Owing to its capacity to enrich the expressivity of the quantum feature space and to the difficulty of simulating it classically, the ZZFeatureMap is one of the most commonly used data encoding methods in quantum classification tasks \cite{ZZKim, ZZAbbas2021Jun, ZZMukhamedieva2024, Vasques2023Jul, Havlicek2019Mar}.

\subsubsection{Challenges and Limitations}

\begin{figure*}[!t]
    \centering
    \includegraphics[width=\textwidth]{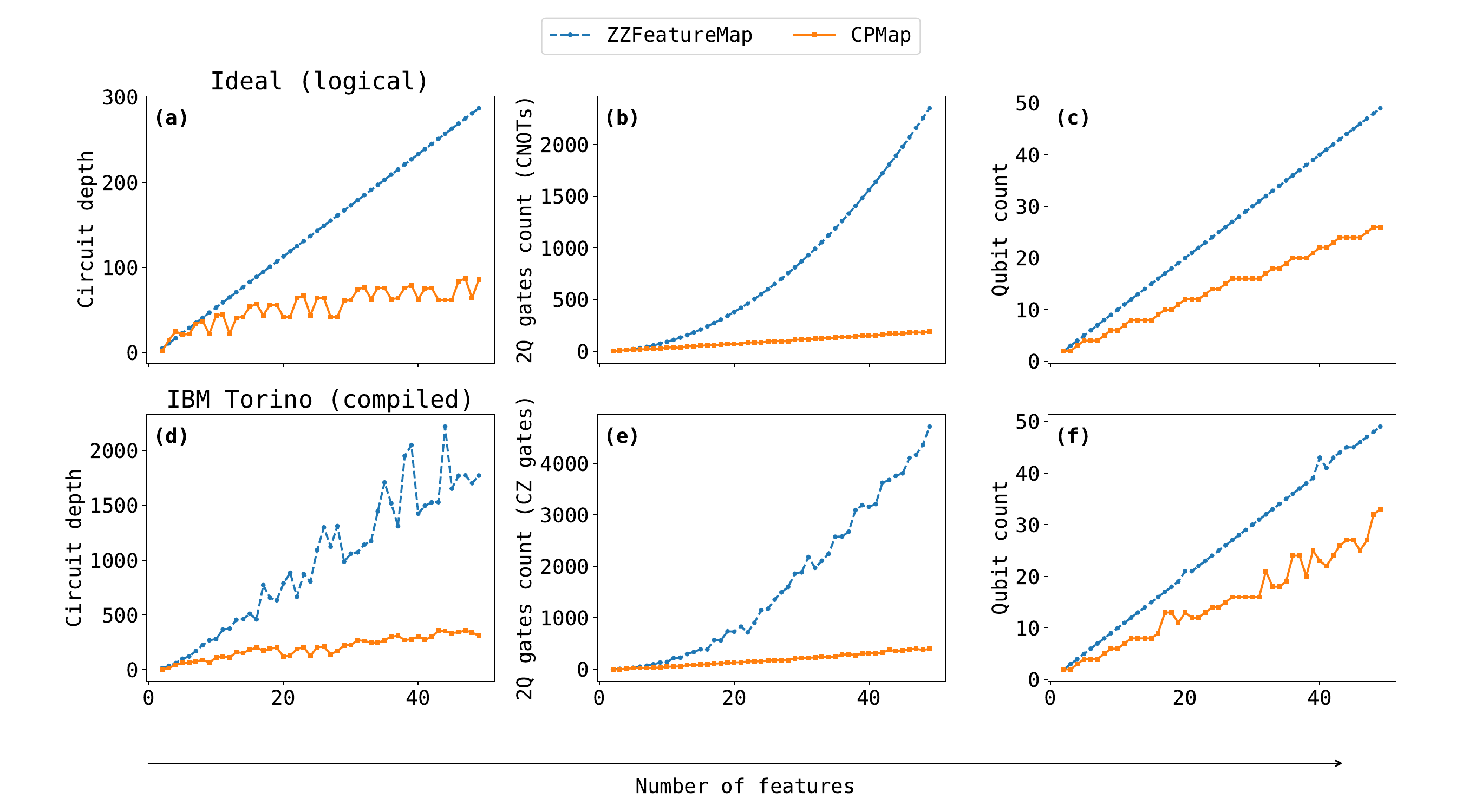}
    \caption{\textbf{Resource scaling of CPMap vs. ZZFeatureMap (ideal vs. hardware-compiled).} The top row reports ideal$/$logical circuit resources (no transpilation) as a function of the number of features: (a) circuit depth, (b) two-qubit gates count (CNOT) operations, and (c) qubit count required by the feature map. The bottom row reports the same quantities after transpiling the circuits to the IBM Torino backend (native gate set and coupling constraints), where the native two-qubit operation is CZ; thus subplot (e) reports CZ counts for the compiled circuits. Across both settings, CPMap exhibits substantially reduced two-qubit resources and depth growth compared to ZZFeatureMap, and the qualitative scaling advantage is preserved after device-aware compilation.}
    \label{fig:resources}
\end{figure*}

All of the aforementioned quantum feature maps require the number of qubits to grow linearly with the dimensionality of the data, requiring one qubit per datum. This requirement becomes prohibitive with high-dimensional data, particularly with current NISQ devices \cite{Preskill2018Aug}. The ZZFeatureMap tends to perform better than the ZFeatureMap so, for this study, we will only consider the ZZFeatureMap. In the case of ZZFeatureMap, the circuit depth grows linearly and the number of CNOT gates in the circuit grows quadratically with the number of features in the data; this will later be seen in Fig.~\ref{fig:resources}.
Appendix~\ref{Appendix_B} explains why we choose the ZZFeatureMap over the ZFeatureMap for our comparisons, as the latter can seldom be used on its own when nonlinear functions of the input data are necessary. Our feature map has better resource costs than the ZZFeatureMap while still performing well, which is crucial for comparison because one can always find feature maps that are less resource intensive but with inferior performance.

\begin{figure*}[!ht]
    \centering
    \includegraphics[scale=1.25]{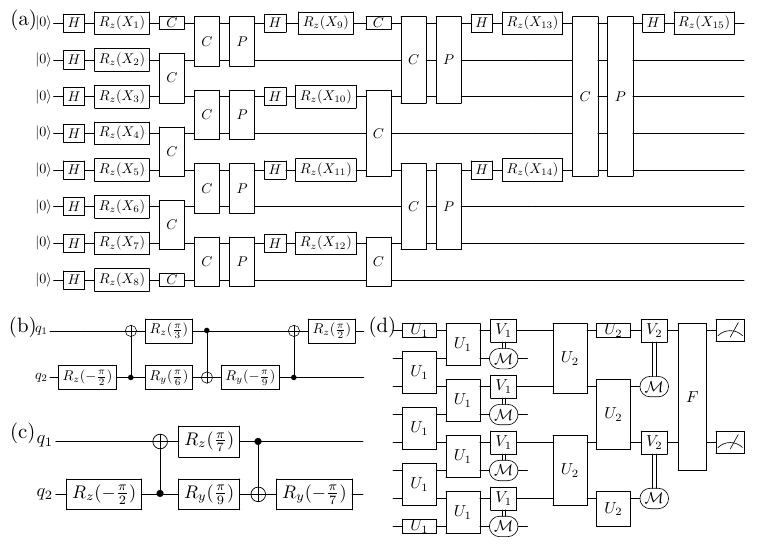}
     \caption{(a) A CPMap with 8 qubits encoding 15 features. The qubits are then output to be used coherently in any application (including measurement, repetition of the kernel, and input to a quantum neural network). $H$ is the Hadamard gate and $R_z(X_i)$ is a single-qubit rotation around the $z$ axis by angle $X_i$. (b) Diagram of the $C$ unitary. (c) Diagram of the $P$ unitary. (d) An illustration of a quantum CNN with 8 qubits. The two-qubit unitaries $U_1$ and $U_2$ perform convolutions, then the pooling operations are done by measurements $\mathcal{M}$ on one qubit followed by feedforward unitaries $V_1$ and $V_2$ on the other qubit; all such operations involve parameters that must be trained. {While (d) uses pooling operations, (a) adopts a similar architectural flow but replaces them with the unitary $P$, allowing more data features to be encoded before the subsequent layer.}}
    \label{fig:intro}
\end{figure*}

\section{Results \label{secIV}}

\subsection{CPMap \label{secIII}}

Our proposed \emph{CPMap} is inspired by the structure of quantum convolutional neural networks (QCNNs) \cite{cong2019quantum}, which themselves borrow ideas from classical convolutional neural networks. Similar to their classical counterparts, QCNNs combine layers that extract salient features from data with pooling operations that progressively reduce dimensionality. This hierarchical reduction allows limited quantum resources to be focused on increasingly abstract representations of the input. 

The CPMap follows a similar guiding principle but adapts it to the data encoding setting. Instead of discarding qubits after a pooling-like operation, the CPMap recycles them to encode additional features. In effect, pooling is replaced by partitioning: qubits freed by earlier feature aggregation are immediately reused for loading new data. Since CPMap is a feature mapping technique used for kernel construction rather than a trainable neural network, parameters are fixed rather than optimized, and measurements occur only at the end. Data are coherently uploaded and processed layer by layer, enabling more features to be embedded on a restricted number of qubits.

The CPMap alternates two types of fixed two-qubit unitaries. The first set, denoted $C$, plays a role analogous to convolutional filters, while the second set, denoted $P$, performs feature focusing similar to pooling. Together, these unitaries reduce the number of active qubits by approximately half at each layer: $n \mapsto n/2$. Iterating this process allows $n$ qubits to encode roughly $2n$ features. More precisely, the maximum number of features that can be encoded on $n$ qubits follows the meta-Fibonacci sequence with parameter $s=0$ \cite{BibEntry2024Jan,BibEntry2024Jan2}:
\begin{equation}
    \begin{split}
        a(N) = & a(N-a(N-1)) + a(N-1-a(N-2)), \\ 
        & \mathrm{with}\quad a(1)=a(2)=1.
    \end{split}
\end{equation}
This construction yields a substantial increase in feature capacity on any given hardware.

The structure of CPMap is illustrated in Fig.~\ref{fig:intro}, which highlights the replacement of QCNN’s measurement-and-feedforward stage with the unitary $P$. More general focusing strategies can be adopted. For instance, if each pooling step reduces $n$ qubits to $n/m$ instead of $n/2$, then $n(m-1)/m$ qubits remain available for subsequent iterations, allowing even more features to be embedded on the same hardware. These flexible design choices make CPMap well suited for encoding large classical datasets into limited quantum registers.

\begin{figure}[H]
\centering   

\includegraphics[scale = 0.8]{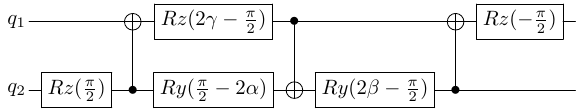}
\caption{A circuit for implementing the three-parameter two-qubit gate $N(\alpha, \beta, \gamma)$, requiring three CNOT gates.}
\label{cnv_circ}
\end{figure}

To instantiate $C$ and $P$, we restrict to two-qubit unitaries of the form
\begin{equation}
    U = (A_{1} \otimes A_{2})\, N(\alpha,\beta,\gamma)\, (A_{3} \otimes A_{4}),
    \label{U4}
\end{equation}
where $A_i\in\mathrm{SU}(2)$ and
\begin{equation}
    N(\alpha,\beta,\gamma) = \exp\!\left[i(\alpha X\otimes X + \beta Y\otimes Y + \gamma Z\otimes Z)\right].
\end{equation}
In general this requires 15 parameters, but in our implementation, we choose $A_i=\mathbb{I}$, leaving only the three free parameters $(\alpha,\beta,\gamma)$. A circuit decomposition of $N(\alpha,\beta,\gamma)$ into three CNOT gates is shown in Fig.~\ref{cnv_circ} \cite{Vatan2004Mar}, and the circuit for the unitaries $C$ and $P$ is shown in Fig.\ref{fig:intro}. 

Let $n$ be the number of physical qubits and $x\in\mathbb{R}^F$ the feature vector. At each layer $\ell$, a subset $S_\ell$ of active qubits is used to encode a contiguous batch of features $\mathbf{x}_\ell$ via $U_Z(\mathbf{x}_\ell)$.

The subset $S_0$ is chosen to be all of the qubits and each subsequent subset $S_\ell$ is chosen to be half of the qubits from the previous subset (e.g. taking every second qubit such that the data encoding takes the hierarchical structure depicted in Fig.~\ref{fig:intro}(a)).

For a pair set $E\subseteq \{\{i,j\}\subset\{1,\dots,n\}: i\neq j\}$ with disjoint pairs, define
\begin{equation}
C[E] \;:=\; \bigotimes_{\{i,j\}\in E} C_{i,j},
\qquad
P[E] \;:=\; \bigotimes_{\{i,j\}\in E} P_{i,j},
\end{equation}
where $C_{i,j}$ and $P_{i,j}$ are the fixed two-qubit unitaries used in the circuit (their internal
decomposition is not important for the formal description).
Even and odd pairings of $S_\ell$ define sets of two-qubit gates $C$ and $P$ (chosen as nearest-neighbours in Fig.~\ref{fig:intro}(a) for convenience on quantum devices). The $\ell$th layer then acts as 
\begin{equation}
\mathcal{L}_\ell(x)=P[E^{(1)}_\ell]\, C[E^{(1)}_\ell]\, C[E^{(0)}_\ell]\, D_{S_\ell}(x_{B_\ell})\, H_{S_\ell}.
\end{equation}
Applying $L$ layers produces the feature map
\begin{equation}
U_{\mathrm{CP}}(x)=\mathcal{L}_L(x)\cdots\mathcal{L}_1(x), \qquad 
\ket{\psi_{\mathrm{CP}}(x)}=U_{\mathrm{CP}}(x)\ket{0}^{\otimes n}.
\end{equation} 

We use this feature map in the fidelity quantum kernel setting, which is given by
\begin{equation}
K_{\mathrm{CP}}(x,x')=\big|\braket{\psi_{\mathrm{CP}}(x)}{\psi_{\mathrm{CP}}(x')}\big|^2.
\label{eqn:11}
\end{equation}

For example, in an 8-qubit circuit we can encode $F=15$ features. The active sets are reduced as $|S_1|=8, |S_2|=4, |S_3|=2, |S_4|=1$. Each layer applies the even/odd $C$ sweeps, and selective $P$, and passes the retained wires forward. This process yields the compact embedding structure shown in Fig.~\ref{fig:intro}.

\paragraph{CNOT count for CPMap:} The CPMap is constructed using the $C$ and the $P$ unitaries in multiple layers ($\ell$), where $C$ has 3 CNOT gates and  $P$ has 2 CNOT gates inside their circuit implementation. At each layer $\ell=0,1,2,\dots$, the active width is given by $|S_\ell|=n/2^\ell$.
And inside each of these layers, we apply $C$ on all active pairs and $P$ on half as many pairs, so the per–layer
CNOT cost is
\begin{equation}
\mathrm{CNOT}_\ell \;=\; 3\cdot \frac{n}{2^\ell}\;+\;2\cdot \frac{n}{2^{\ell+1}}
\;=\;\frac{4n}{2^\ell}.
\end{equation}

Summing $L$ full layers gives
\begin{equation}
 \mathrm{CNOT}_{\le L} \;=\; \sum_{\ell=0}^{L-1}\frac{4n}{2^\ell}
\;=\; 8n\bigl(1-2^{-L}\bigr).   
\end{equation}
In terms of the number of features encoded after $L$ layers,
\begin{equation}
F_L \;=\; \sum_{\ell=0}^{L-1}\frac{n}{2^\ell}
\;=\; 2n\bigl(1-2^{-L}\bigr),
\end{equation}
so the CNOT count can be written as
\begin{equation}
    \boxed{\;\mathrm{CNOT}_{\le L} \;=\; 4\,F_L\;}
    \label{eq15}
\end{equation}
and, in the limit $L\to\infty$ (i.e., $F\!\to 2n$), $\mathrm{CNOT}_{\mathrm{total}} \to 8n = 4F$.

\subsubsection{Resource Complexity}
Encoding a data point with $F$ features using the ZZFeatureMap typically requires $n=F$ qubits and approximately $\mathcal{O}(F^2)$ two-qubit entangling operations, leading to rapidly increasing depth as $F$ grows. In contrast, CPMap encodes $F$ features using substantially fewer qubits, with $n \approx F/2$ (cf. Eq.~\eqref{eq15}), and exhibits an approximately linear growth in entangling cost, i.e., $\mathcal{O}(F)$ two-qubit operations in the ideal (pre-compilation) circuit. Figure~\ref{fig:resources} reports this comparison in two settings: (top row) the logical circuits prior to transpilation, where the two-qubit count is reported in terms of CX (CNOT) gates, and (bottom row) the same circuits after transpilation to the IBM Torino backend, where the native two-qubit operation is CZ and the compiled two-qubit cost is therefore reported as CZ count. Importantly, the qualitative resource advantage of CPMap (lower depth growth, fewer two-qubit gates, and reduced qubit requirements) is preserved under device-aware compilation, making it particularly attractive for near-term (NISQ) implementations.

For completeness, we provide a hardware-feasibility analysis based on backend-timing circuit durations and device coherence times ($T_1$, $T_2$) in Appendix~\ref{app:coherence_runtime} (see Fig.~\ref{fig:coherence}). This shows that the compiled gate schedule remains well within coherence across the tested feature dimensions.

\subsection{Numerical Results \label{secIV}}

The performance of the CPMap was benchmarked against several established quantum and classical kernels to ascertain its capabilities. We use several datasets characterized in Table~\ref{table2} to check its performance as an SVC for classifying data into discrete categories. \\

\begin{table*}[ht!]
\centering
\begin{tabular}{|l|l|l|l|}
\hline
\textbf{Dataset} & \textbf{Number of Features} & \textbf{Number of Instances} & \textbf{Classes} \\
\hline
\href{https://www.kaggle.com/datasets/jamieleech/ionosphere}{Ionosphere} & 34 & 351 & 2\\
\hline
\href{https://archive.ics.uci.edu/dataset/17/breast+cancer+wisconsin+diagnostic}{Breast cancer Diagnostic (Wisconsin)} & 30 & 569 & 2\\
\hline
\href{https://www.kaggle.com/datasets/mlg-ulb/creditcardfraud}{Credit card Fraud detection
} (Balanced) & 30 & 984 & 2 \\
\hline
\href{https://archive.ics.uci.edu/dataset/174/parkinsons}{Parkinson's disease (PD)} & 22 & 195 & 2 \\
\hline
\href{https://www.kaggle.com/datasets/fedesoriano/stellar-classification-dataset-sdss17}{Stellar Classification Dataset - SDSS17} & 17 & $100000^{*}$ & $3^{*}$ \\
\hline
\href{http://archive.ics.uci.edu/dataset/45/heart+disease}{Heart Disease} & 11 & 918 & 2 \\
\hline
\href{https://www.kaggle.com/datasets/yasserh/titanic-dataset}{Titanic} & 10 & 891 & 2 \\
\hline
\href{https://www.kaggle.com/datasets/uciml/mushroom-classification}{Agaricus Lepiota Mushrooms} & 21 & 8145 & 2 \\
\hline
\end{tabular}
\caption{Properties of various datasets used in this study. *The Stellar Classification Dataset properly has three classes, but we used only two of them, Galaxy and Star, combining for a total of 81039 data points.}
\label{table2}
\end{table*}

Owing to the skewed nature of the majority of our test data, we opted for the Matthews Correlation Coefficient (MCC) as our primary evaluation criterion. For an in-depth explanation, kindly refer to Methods and Appendix~\ref{Appendix_A}. MCC serves as a reliable metric, offering a comprehensive assessment of binary classification outcomes by considering both true and false positives as well as negatives. A score of +1 in MCC denotes perfect prediction accuracy, 0 suggests no better than a random prediction, and -1 signifies a complete mismatch between the predicted and actual outcomes. 
It overcomes metrics such as accuracy that can give misleadingly positive results when classifiers are tested on skewed data.

In our research, we extensively utilized Qiskit's Statevector simulator, a tool designed to simulate the ideal quantum states of a quantum system without any external noise or decoherence. This simulator provides a precise representation of the quantum state, allowing for accurate computations and predictions. It is particularly beneficial for theoretical explorations and understanding the ideal behaviour of quantum algorithms. Alongside the Statevector simulator, we also employed Qiskit's noisy simulators. These are designed to mimic real IBM quantum devices by leveraging system snapshots. Such snapshots capture vital data about the quantum setup, including the coupling map, foundational gates, and qubit attributes (T1, T2, error rates, and more), proving instrumental for transpiler testing and conducting system simulations with noise. These noisy simulators present a more grounded view of quantum operations in tangible settings. Their use enabled us to assess the durability of our algorithms in authentic environments and refine them to better withstand quantum disruptions and other unforeseen challenges. We then proceeded to run small trials on ibm\_quebec and ibm\_torino.

\subsubsection{Results with Seven Features After PCA}
In the initial phase of our study, we employed principal component analysis (PCA) to reduce the feature set size to seven across all datasets. Specifically, we limited the sample size of the Stellar Classification Dataset to 2,000 and excluded the QSO class to expedite the simulation process. For support vector classification, we leveraged the capabilities of both the SciKit Learn~\cite{scikit-learn} and QisKit~\cite{Qiskit} libraries.

\begin{figure}[!t]
    \centering
    \includegraphics[width=0.5\textwidth]{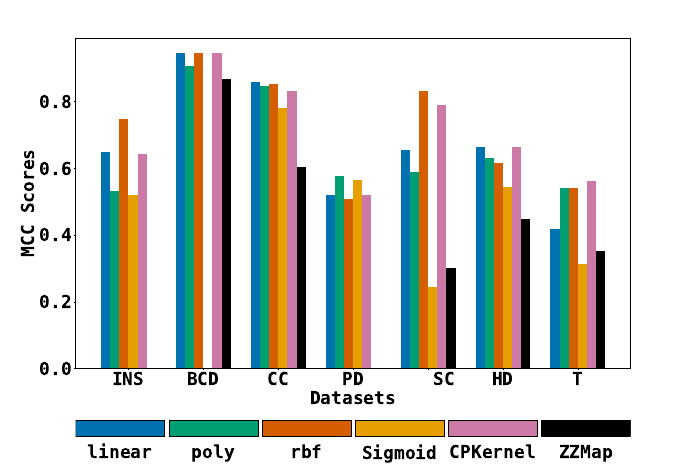}
    \caption{MCC-score-based analysis of different kernels on different datasets scaled to each have 7 features, comparing the effectiveness of each kernel in handling specific types of data. The CPMap (second from right for each dataset) outperforms the ZZFeatureMap (rightmost) on all datasets in this 7-feature benchmark, and is competitive with strong classical kernels (linear, poly, RBF, sigmoid), sometimes exceeding them depending on the dataset.    
    For some datasets, certain kernels completely failed, so there is no bar visible for the ZZFeatureMap for the INS and PD datasets and for the sigmoid kernel for the BCD dataset. Dataset acronyms: Ionosphere (INS), Breast Cancer Diagnostic (BCD), Credit Card Fraud (CC), Parkinson's Disease (PD), Stellar Classification (SC), Heart Disease (HD), and Titanic Survival (T).
}
    \label{mcc7}
\end{figure}

Our findings, depicted in Fig.~\ref{mcc7}, reveal that the CPMap consistently outshines the ZZFeatureMap quantum kernel across a diverse range of datasets. Remarkably, there were instances where the CPMap not only matched but exceeded the performance of standard classical kernels [linear, polynomial (poly), radial basis function (RBF), sigmoid], highlighting its promising applications in the realm of quantum machine learning. The accompanying plots, which display MCC scores, further substantiate the superior efficacy of the CPMap. While some classical kernels exhibited strong performance on specific datasets, they faltered on others. In contrast, the CPMap consistently delivered robust results, often rivaling or surpassing the best-performing classical kernels.

For some datasets, we repeated the CPMap up to two times but uploaded the same features for each repetition. This increases the total number of CNOT gates by a factor of two but maintains the total number of qubits and the significant reduction in the number of CNOT gates; see Appendix~\ref{app:repetitions 7} for details. An additional multi-class experiment on the Stellar Classification dataset (balanced $N=1500$) is reported in Appendix~\ref{multiclass}, and an expanded comparison against fixed data re-uploading feature-map variants (DR1--DR6) is provided in Appendix~\ref{data_reuploading_fm}.

\subsubsection{Results for the Noisy Simulation with Seven Features}
\begin{figure}[!t]
    \centering
    \hspace*{0cm}   
    \includegraphics[width=0.5\textwidth]{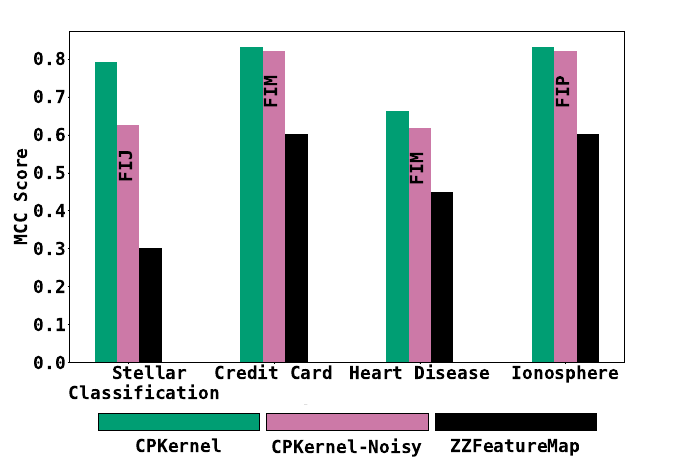}
    \caption{MCC score for the noisy simulation of CPMap on different datasets. Here we used \textit{'fake\_ibmq\_jakarta' (FIJ)}, \textit{'fake\_ibmq\_manila' (FIM)} and \textit{'fake\_ibmq\_perth' (FIP)} backends.}
    \label{mcc_noisy}
\end{figure}
In addition to standard simulations, we conducted noisy simulations on three specific datasets: Stellar Classification, Balanced Credit Card, and Heart Disease.

As demonstrated in Figure \ref{mcc_noisy}, CPMap remains competitive under these noise models and, in this benchmark, achieves higher MCC than ZZFeatureMap even when ZZFeatureMap is evaluated in the ideal (noiseless) setting.

\subsubsection{Results for the Noisy Simulation with sixteen Features}
\begin{figure}[!t]
    \centering
    \includegraphics[width=\linewidth]{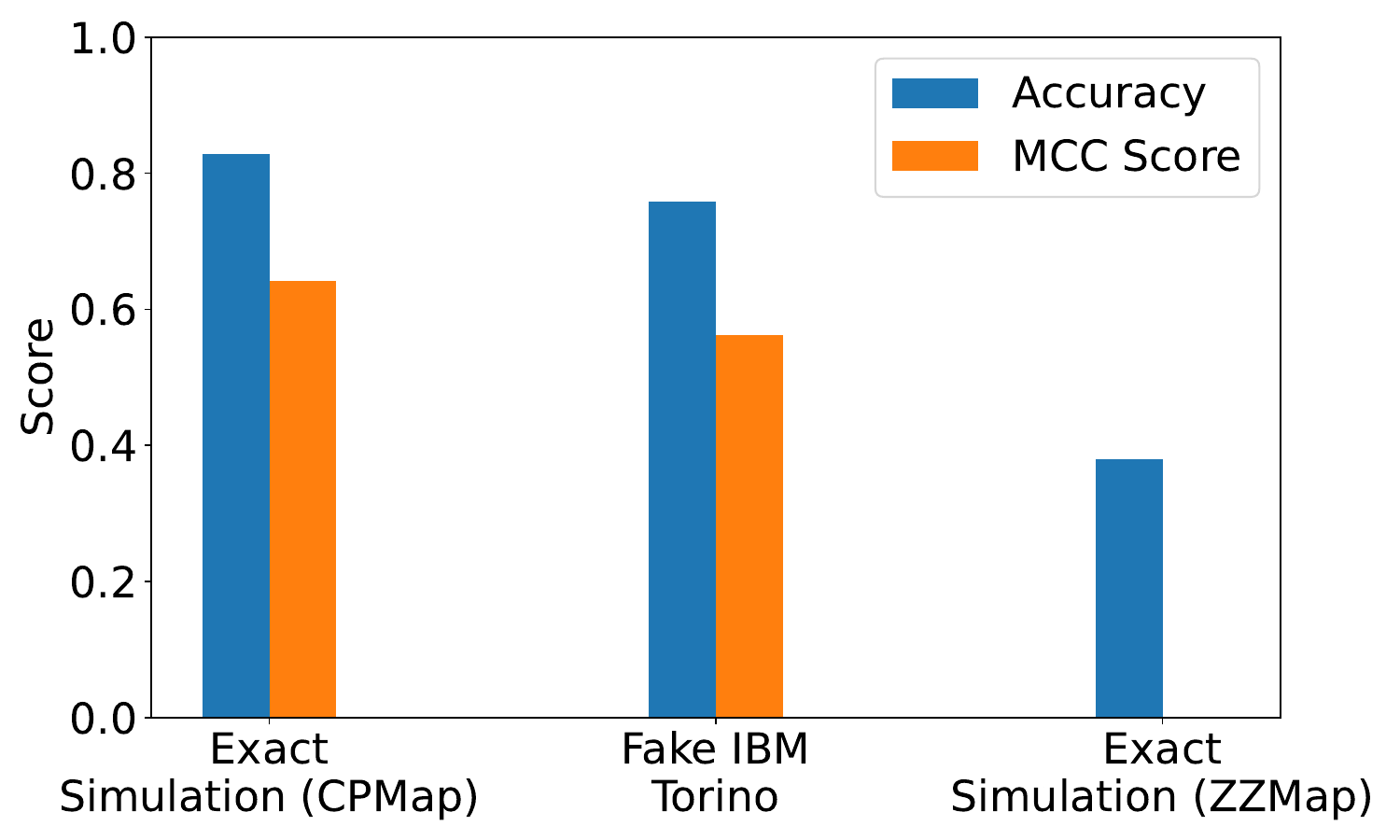}
    \caption{Performance comparison of the CPMap under ideal and noisy conditions. The CPMap maintains reasonable classification performance despite noise, achieving an MCC of $0.56$ and accuracy of $0.76$ in the noisy case, compared to $0.64$ and $0.83$, respectively, in the ideal case. While the ZZMap achieves $0$ MCC. These results highlight the noise resilience of the CPMap encoding strategy.
}
    \label{fig:park_noisy}
\end{figure}

To check the effect of realistic noise when dealing with a higher-dimensional problem, we ran a similar noise test using the Parkinson's disease dataset using the Qiskit fake back-end model of $\texttt{ibm\_torino}$. This time, we used the dataset with 16 features (i.e., 9 qubits for CPMap). The results are shown in figure~\ref{fig:park_noisy}.

The CPmap under ideal conditions achieved an accuracy of $0.83$ and MCC of $0.64$, while under noisy conditions, the performance degraded, but it still achieved an accuracy of $0.76$ and MCC of $0.56$, highlighting its resilience to the realistic noisy conditions of current devices. The ZZFeatureMap failed to classify the dataset, achieving an MCC of 0. 

\subsubsection{A case for quantum usefulness}
\begin{figure}[!t]
    \centering
    \includegraphics[width=0.525\textwidth]{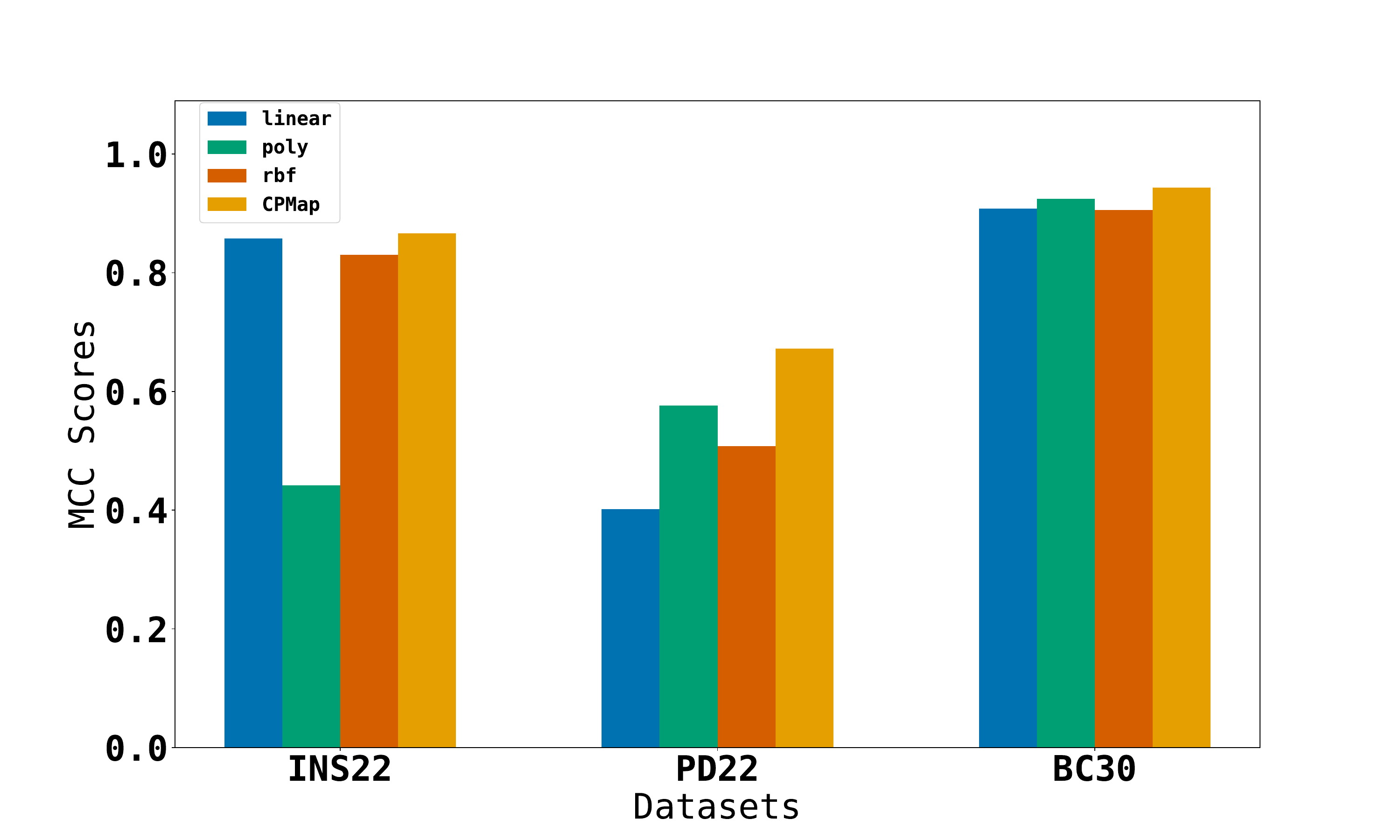}
    \caption{This bar chart illustrates the MCC scores achieved by four kernel functions—from left to right: linear, polynomial (poly), radial basis function (RBF), and the CPMap—when applied to three different datasets.}
    \label{mcc_adv}
\end{figure}

Next, we turn to datasets with so many features that it is prohibitive to simulate the ZZFeatureMap on a classical computer due to the exorbitant qubit and gate-count requirements. We perform PCA on the Ionosphere dataset to reduce it to 22 features, in order to analyze it with the exact same CPMap as for the Parkinson's dataset that has exactly 22 features.

The bar chart presented in Fig.~\ref{mcc_adv} illustrates the performance of various kernels, including linear, poly, RBF, and CPMap, across the three distinct datasets: Ionosphere (INS22), Parkinson's Disease (PD22), and Breast Cancer (BC30). Notably, the CPMap outperforms its counterparts across all datasets, achieving the highest MCC scores, indicative of its superior predictive capabilities. On the PD22 dataset, CPMap's score is a significant leap from poly's score and is even greater compared to the linear and RBF kernels. This trend persists with the INS22 and BC30 datasets, where CPMap consistently maintains the lead. Similar results are presented in Appendix~\ref{app:full} with even more features per dataset and results for different numbers of repetitions of the CPMap in Appendix~\ref{app:repetitions full}.

The consistent outperformance of CPMap in our experiments indicates an empirical advantage for this embedding under the studied settings, particularly when feature counts are larger and circuit resources are constrained. The substantial margin by which CPMap leads for datasets with large numbers of features hints at its unique ability to capture complex patterns that classical kernels might not discern as effectively, thereby bolstering the hypothesis that quantum machine learning could offer computational benefits over traditional algorithms.

\subsection{Statistical Significance Testing} 
In addition to demonstrating quantum usefulness, we further validated the performance of our custom CPMap kernel against traditional RBF, poly, and linear kernels using Parkinson's disease dataset. We employed statistical significance testing to confirm the superiority of CPMap in terms of accuracy and the MCC. \\

We performed paired t-tests to evaluate the significance of performance differences between the CPMap kernel and the traditional kernels (RBF, Poly, and Linear) in terms of accuracy and MCC scores across 80 independent runs. The results are summarized in Table~\ref{t_test_results}.

\begin{table}[h!]
    \centering
    \caption{Paired t-Test Results Comparing CPMap with Other Kernels}
    \label{t_test_results}
    \begin{tabular}{lcccc}
        \hline
        \textbf{Comparison} & \textbf{Metric} & \textbf{t-Statistic} & \textbf{p-Value} \\
        \hline
        RBF vs. CPMap & Accuracy & 8.45 & 1.15 $\times$ 10$^{-12}$ \\
        RBF vs. CPMap & MCC & 8.54 & 7.83 $\times$ 10$^{-13}$ \\
        \hline
        POLY vs. CPMap & Accuracy & 11.38 & 2.57 $\times$ 10$^{-18}$ \\
        POLY vs. CPMap & MCC & 11.32 & 3.28 $\times$ 10$^{-18}$ \\
        \hline
        Linear vs. CPMap & Accuracy & 11.60 & 9.78 $\times$ 10$^{-19}$ \\
        Linear vs. CPMap & MCC & 11.26 & 4.26 $\times$ 10$^{-18}$ \\
        \hline
    \end{tabular}
\end{table}

The t-statistics and p-values indicate that the performance differences between the CPMap kernel and each of the traditional kernels are statistically significant. The t-statistics exceed typical critical values, and the p-values are significantly below standard significance levels (e.g., 0.05, 0.01), thereby rejecting the null hypothesis of no difference in performance.

\paragraph{Performance Metrics Summary:}

To provide a detailed view of the CPMap kernel's performance, we calculated the mean and standard deviation of accuracy and MCC scores across the 80 runs for each kernel. These metrics are detailed in Table~\ref{mean_std_accuracy_mcc}.

\begin{figure}[!t]
    \centering
    \includegraphics[width=0.5\textwidth]{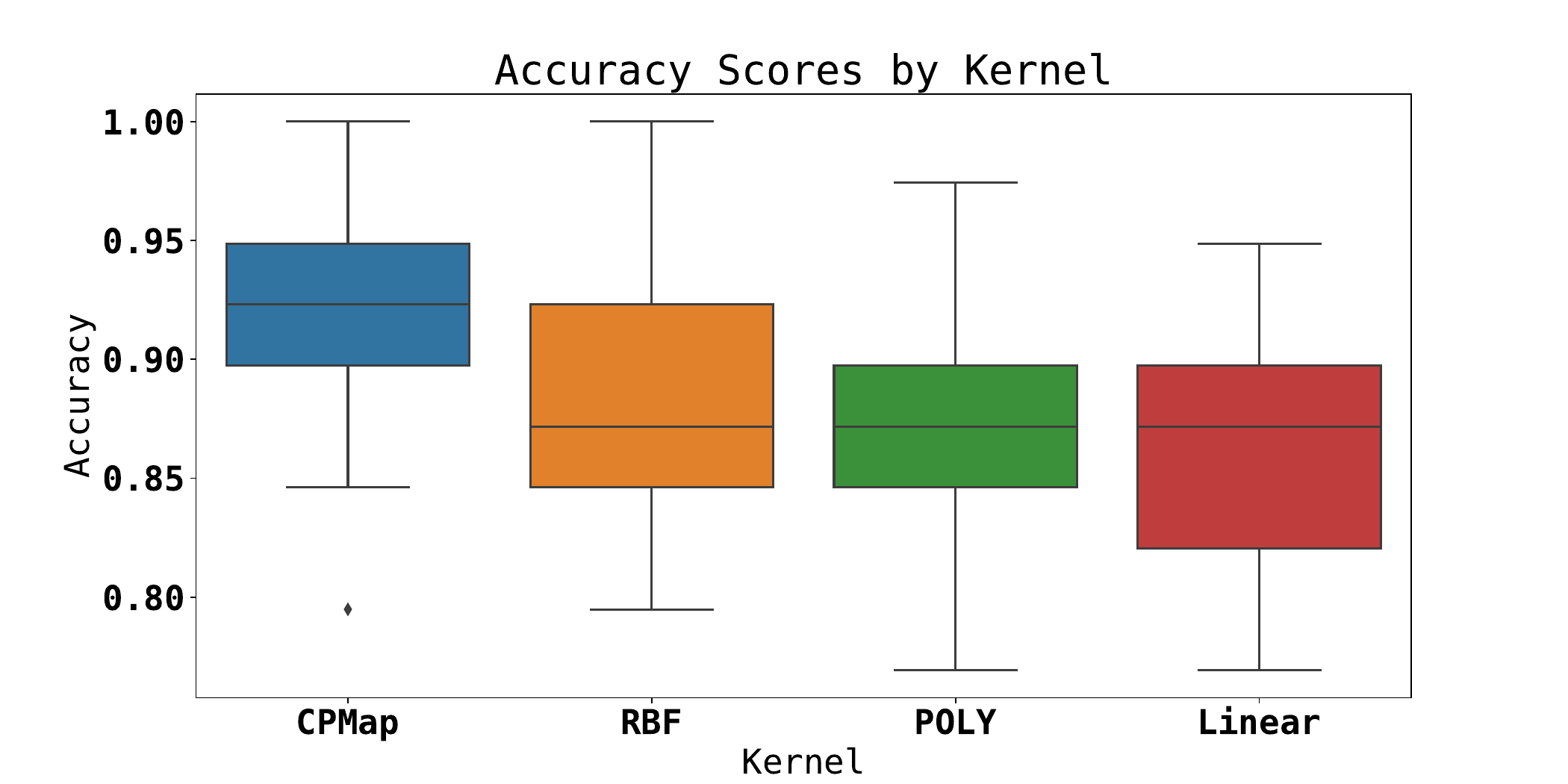}
    \caption{Box Plot of Accuracy Scores for Different Kernels}
    \label{acc_boxplot}
\end{figure}

\begin{figure}[!t]
    \centering
    \includegraphics[width=0.5\textwidth]{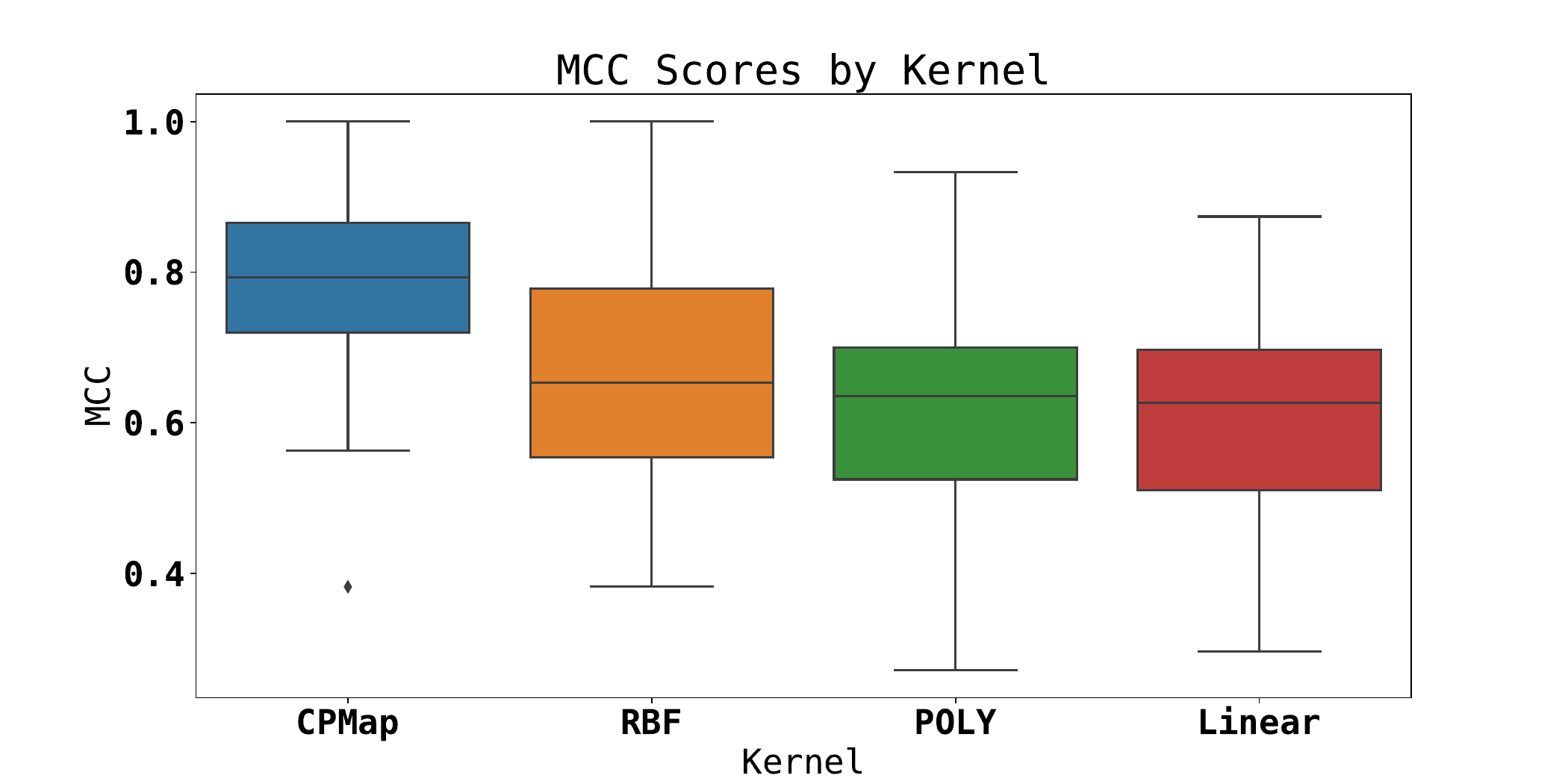}
    \caption{Box Plot of MCC Scores for Different Kernels}
    \label{mcc_boxplot}
\end{figure}

\begin{table}[h!]
    \centering
    \caption{Mean and Standard Deviation of Accuracy and MCC for Different Kernels}
    \label{mean_std_accuracy_mcc}
    \begin{tabular}{lcccc}
        \hline
        \textbf{Kernel} & \textbf{Accuracy} & \textbf{Accuracy} & \textbf{MCC} & \textbf{MCC} \\
        \textbf{} & \textbf{Mean} & \textbf{Std} & \textbf{Mean} & \textbf{Std} \\
        \hline
        Linear       & 0.8615 & 0.0442 & 0.6102 & 0.1333 \\
        Poly         & 0.8660 & 0.0424 & 0.6135 & 0.1354 \\
        CPMap  & 0.9234 & 0.0388 & 0.7898 & 0.1118 \\
        RBF          & 0.8785 & 0.0477 & 0.6532 & 0.1438 \\
        \hline
    \end{tabular}
\end{table}
Figures~\ref{acc_boxplot} and~\ref{mcc_boxplot} illustrate the distribution of accuracy and MCC scores across the different kernels. These visualizations highlight the consistent superior performance of the CPMap kernel.

This statistical analysis and visualization confirm the significant advantage of the CPMap kernel over traditional kernels in the context of Parkinson's disease data, reinforcing the potential for quantum machine learning to achieve empirical improvement in predictive tasks. 

\subsection{Test on IBM's Hardware}

{To investigate the feasibility and performance of our CPMap quantum kernel on real superconducting quantum processors, we conducted experiments on two IBM devices: the 127-qubit ibm\_quebec (Eagle-class) and the 133-qubit ibm\_torino (Heron-class). All experiments targeted a binary classification task using the Parkinson’s disease dataset, which originally contains 195 samples and 22 numerical features per instance. The qubit configurations, data preprocessing, and mitigation strategies can be found in the Methods section.}

\subsubsection{Test with full dataset on ibm\_quebec}
We evaluated the CPMap kernel on IBM’s 127-qubit ibm\_quebec device using the full 22-feature Parkinson dataset encoded on 12 qubits. To assess kernel performance, we ran tests across three transpiler optimization levels and applied error mitigation techniques including dynamical decoupling (DD), zero noise extrapolation (ZNE), and probabilistic error cancellation (PEC). In all cases, the average entropy of the output was approximately same as the number of qubits, which implies the outputs yielded no meaningful information. Specifically, entropy values ranged from 11.58 to 11.66 across all runs—very close to the maximum possible for a 12-qubit system. This was attributed both to CPMap’s requirement for all-to-all connectivity—which introduces numerous SWAP operations and deepens the circuit—and to the high noise levels inherent to the ibm\_quebec device.

We also performed a small test on the newer 133-qubit ibm\_torino (Heron-class) device using a subset of the Parkinson dataset. In contrast to ibm\_quebec, the output entropy on ibm\_torino remained within a reasonable range, with values ranging from 5.59 to 9.79—well below the 12-qubit maximum—indicating that the circuits retained meaningful structure. This suggests that the full dataset could feasibly be processed on this hardware. However, due to limited device access, we conducted the experiment with 7 PCA-reduced features mapped to 4 qubits. The classification results are reported below.

\subsubsection{Test with PCA-reduced dataset on ibm\_quebec}

\begin{figure}[ht]
    \centering
    \includegraphics[width=0.5\textwidth]{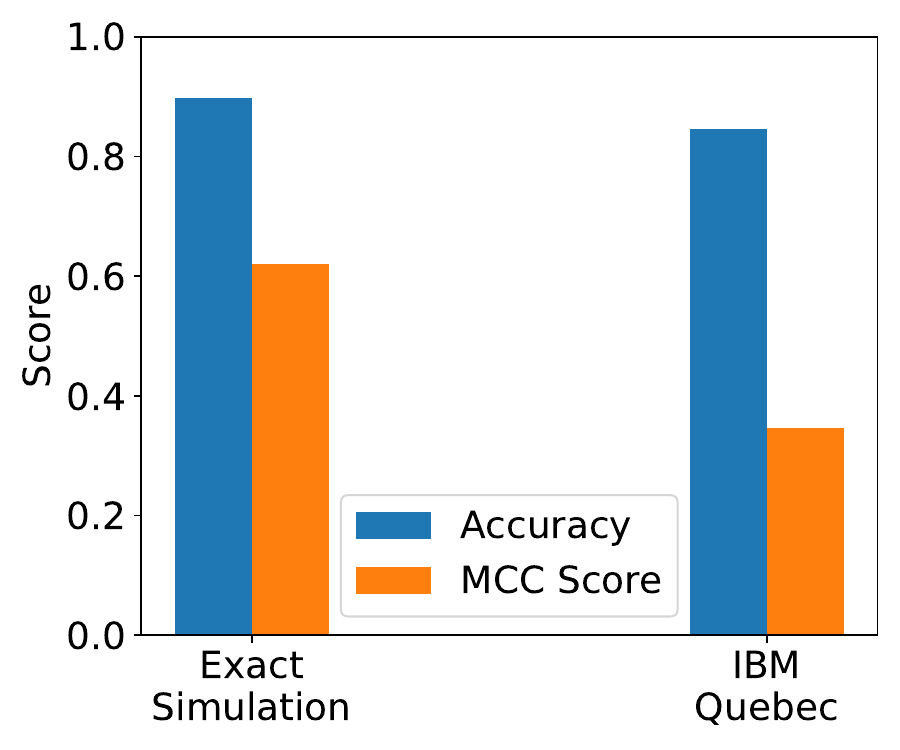}
    \caption{Performance comparison of the CPMap kernel for Parkinson’s disease classification. Accuracy and MCC Score are shown for both exact simulation and IBM's ibm\_quebec (4-qubit encoding).}
    \label{ibm_comparison_bar}
\end{figure}

We tested the CPMap kernel on IBM’s ibm\_quebec device using 4 qubits (7 input features) and no error mitigation for Parkinson’s disease classification. Experiments were performed on qubit chains 39-40-41-42 or 39-40-41-53. The calibration data for these qubits indicate the following average values: $3.39\times 10^{-3}$ s for T1, $2.37\times 10^{-3}$ s for T2, and $3.82\times 10^{-3}$ for the two-qubit gate error rate. In this trial, we obtained an MCC score of $0.34$ and an accuracy comparable to simulation (see Fig~\ref{ibm_comparison_bar}). While the result is still affected by hardware noise, it marks the first instance where this kernel produced a meaningful result on real hardware.

As mentioned above, one limiting factor is the device’s restricted qubit connectivity, which requires the transpiler to insert multiple SWAP gates during compilation, increasing the number of CNOT operations and overall circuit depth. This opens the path to explore optimal compilation of our kernel in architectures with limited connectivity and realization in platforms such as trapped ions where qubit connectivity is not constrained.

\subsubsection{Test with PCA reduced dataset on ibm\_torino}

In this case, we deployed our CPMap kernel circuit on IBM’s ibm\_torino, a 133-qubit superconducting quantum processor belonging to the Heron family. Heron-class devices feature improvements in qubit coherence and native two-qubit gate implementations. At the time of execution, the backend-reported median coherence times were approximately $T_1 \approx 1.83\times 10^{-4}\,\mathrm{s}$ and $T_2 \approx 1.41\times 10^{-4}\,\mathrm{s}$.

Due to limited hardware access, we performed this experiment using a PCA-reduced representation with $d=7$ features, encoded on 4 qubits using CPMap. The overlap circuits required for the fidelity kernel were executed on hardware using IBM Runtime’s \texttt{Sampler} primitive with otherwise default runtime settings. We used 1024 shots per circuit and did not apply any error mitigation or correction techniques for this result. We did not manually select a qubit layout; instead, we allowed the IBM Runtime compilation workflow to choose the physical qubits and initial layout automatically. All circuits were transpiled with optimization level 3.

The quantum kernel matrix was computed in a single run across the full dataset 
and training and test matrices were extracted without recomputation. We allocated 80 samples for training and 20 for testing.

The quantum kernel matrix was computed in a single run (i.e., without recomputing entries across training and test evaluation), and the training and test kernel submatrices were extracted from this precomputed kernel. We allocated 80 samples for training and 20 samples for testing.

\begin{figure}[H]
    \centering
    \includegraphics[width=0.47\textwidth]{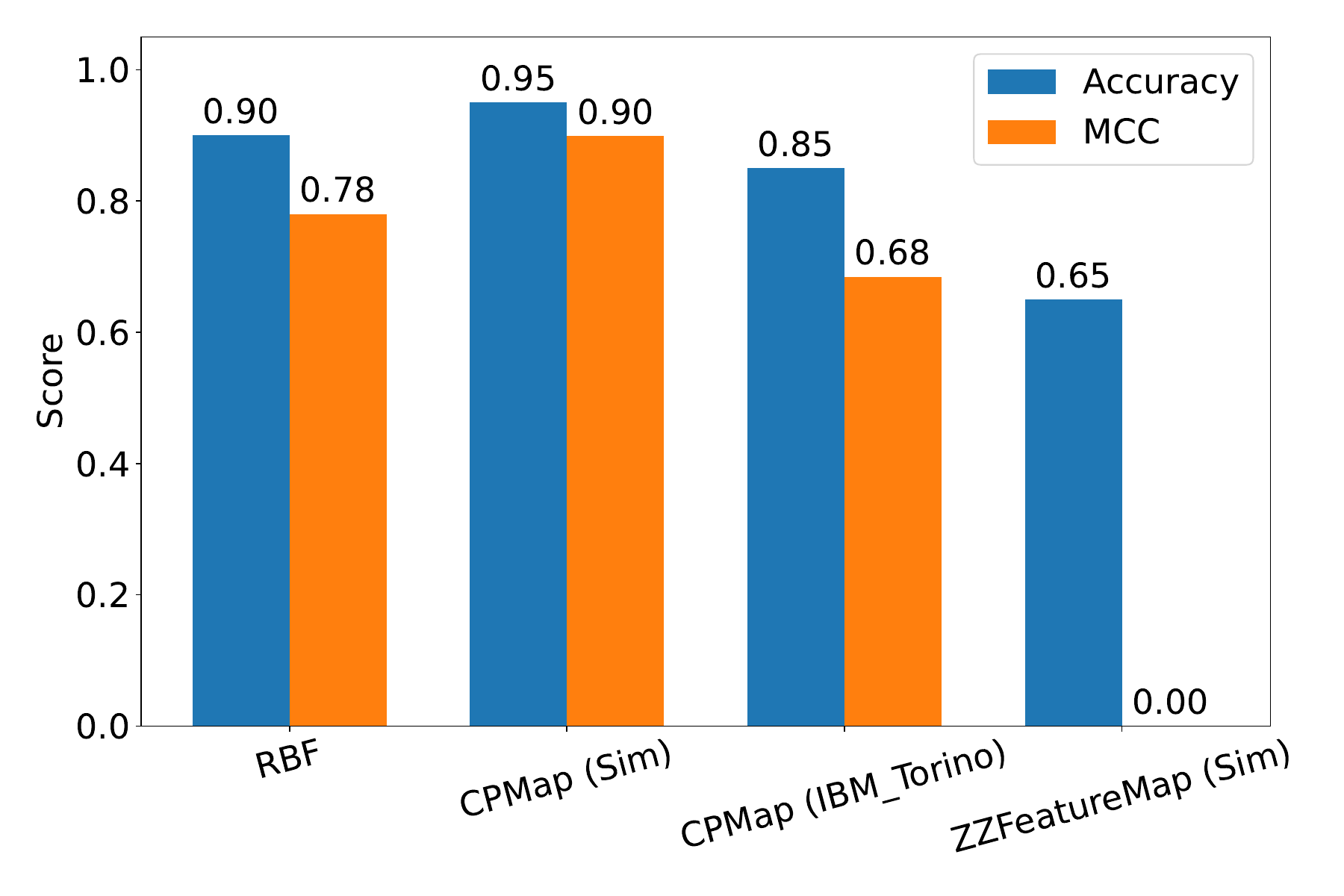}
    \caption{Comparison of CPMap based kernel on the curated Parkinson dataset with PCA-reduced input ($d=7$), executed on \texttt{ibm\_torino} (Sampler, 1024 shots, optimization level 3; no error mitigation). The simulated CPMap-based Kernel achieves the highest performance, followed by RBF and quantum hardware (ibm\_torino). }
    \label{ibm_torino_comparison_bar}
\end{figure}

Despite inevitable hardware noise, the quantum kernel executed on ibm\_torino demonstrated strong performance. It achieved an MCC of 0.68 and an accuracy of 0.85 (see Fig.~\ref{ibm_torino_comparison_bar}). Among the three kernels evaluated, the simulated CPMap kernel achieved the highest performance (MCC of 0.90, accuracy of 0.95), followed by the classically optimized RBF kernel (MCC of 0.78, accuracy of 0.90), and finally, the CPMap executed on real hardware. Notably, the ZZFeatureMap kernel, simulated under \textit{ideal noiseless} conditions, yielded an MCC of 0.00, failing to capture any useful structure in this task. This further underscores the practical relevance of our CPMap. 

The observed performance gap between the statevector simulation and the real-device execution can be attributed to decoherence time, gate and measurement errors in the compiled overlap circuits. Nevertheless, the successful deployment of CPMap on ibm\_torino confirms its feasibility on today’s NISQ devices and underscores the promise of quantum kernels in practical machine learning tasks.


\subsection{Parameter sensitivity and kernel geometry}
\begin{figure}[ht]
    \centering
    \includegraphics[width=0.47\textwidth]{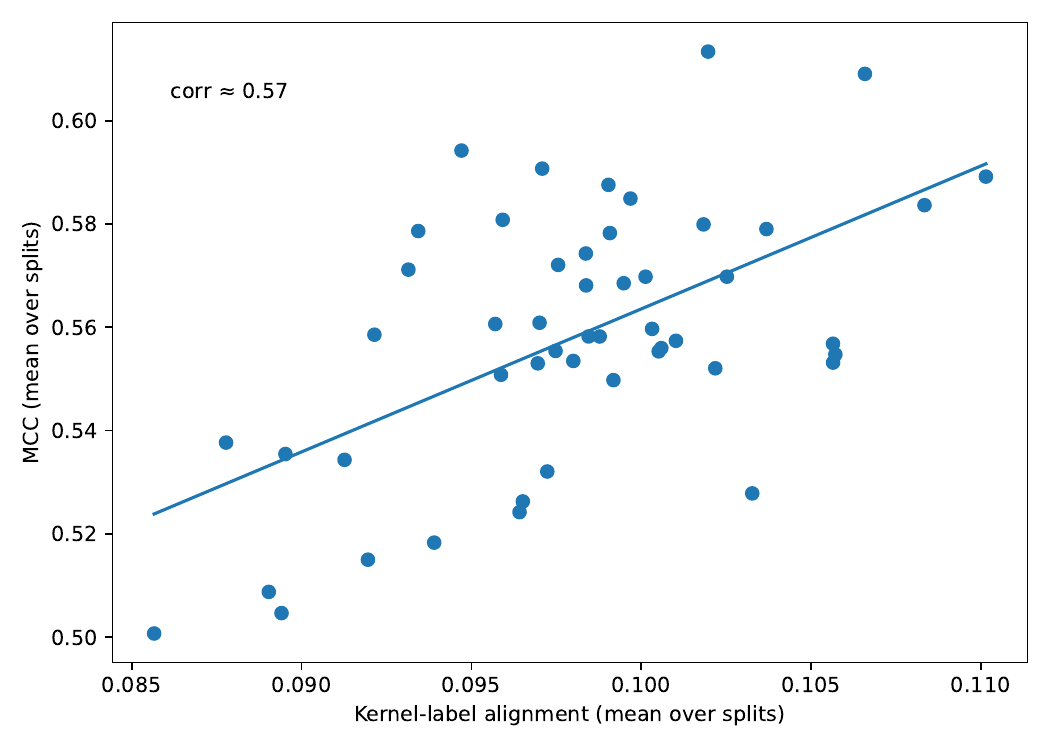}
    \caption{Kernel geometry vs performance under random parameter search on Breast Cancer ($n=400$, $d=12$). Each point is one parameter setting evaluated over 50 stratified splits. The x-axis is kernel–label alignment (higher means the kernel similarities better match class structure) and the y-axis is mean MCC. A strong positive trend (corr $\approx 0.57$) shows that CPMap parameters systematically shape kernel geometry in a way that predicts performance. The wide dispersion in MCC further shows that the parameters materially affect results. The solid line is a least-squares fit.}
    \label{kernel_alignment}
\end{figure}

CPMap contains a small set of fixed design parameters that determine how pairs of qubits interact inside the feature map. These parameters are not trained, so it is important to check whether the method is stable and whether parameter variations induce meaningful (rather than arbitrary) changes in the resulting kernel. To do this, we performed a random parameter search on a 400-sample subset of the Breast Cancer dataset: we randomly sampled CPMap angle settings (varying both the C and P-block parameters), constructed the corresponding kernel matrices, and evaluated the same SVM pipeline using the mean performance over 50 stratified train$/$test splits. For each sampled setting, we also computed kernel–label alignment, a standard scalar summary of how well the kernel similarity matrix matches the class structure (higher alignment means that samples from the same class tend to be more similar under the kernel than samples from different classes). As shown in Fig. \ref{kernel_alignment}, parameter settings with higher alignment consistently yield higher classification performance (correlation $\approx$ 0.57). This indicates that the CPMap parameters control the geometry of the induced kernel in a systematic way, rather than acting as arbitrary constants, and it provides a practical diagnostic for selecting reasonable parameter regimes. Another point to note is that the performance varies substantially across sampled settings (a wide spread in MCC), demonstrating that the CP parameters are not innocuous constants; rather, they can materially influence the quality of the kernel and downstream accuracy. Taken together, these results support two practical conclusions: (i) CPMap is not overly brittle---many parameterizations remain competitive---but (ii) the choice of angles does matter, and alignment provides a simple diagnostic for identifying promising regimes without introducing trainable parameters. Additional experimental details and kernel diagnostics are provided in Supplementary Section~\ref{hyperparam}.

\subsection{Issue of Exponential Concentration:}

Exponential concentration of quantum kernels, as formalized in \cite{thanasilp2024}, presents a fundamental limitation for fidelity-based kernels. Given a quantum feature map $U(x)$ on $n$ qubits, the fidelity kernel is defined as
\begin{equation}
k(x,x') \;=\; \big|\langle 0^n | U(x)^\dagger U(x') | 0^n \rangle \big|^2 \;=\; p_{0}(x,x'),
\end{equation}
and the variance of $p_{0}$ across random pairs $(x,x')$ quantifies the kernel’s discriminative spread. As the feature dimension increases, the overlaps concentrate around their mean, leading to
\begin{equation}
\mathrm{Var}[p_{0}] \;\sim\; e^{-c \cdot n}, \qquad c>0,
\end{equation}
so that kernels become progressively uninformative in high dimensions.

Our proposed CPMap and its implementation based on the fidelity kernel, by construction, is not exempt from the issue of exponential concentration. However, the CPMap changes the rate of the exponential concentration by quadratically lowering the CNOT counts and halving the qubit overhead. In addition to the effects of resource efficiency on slowing the exponential concentration, we observed that CPMap behaves more favorably as a function of the number of qubits, thereby providing a more realistic paradigm for information processing on NISQ devices.

\begin{figure}[H]
    \centering
    \includegraphics[width=0.9\linewidth]{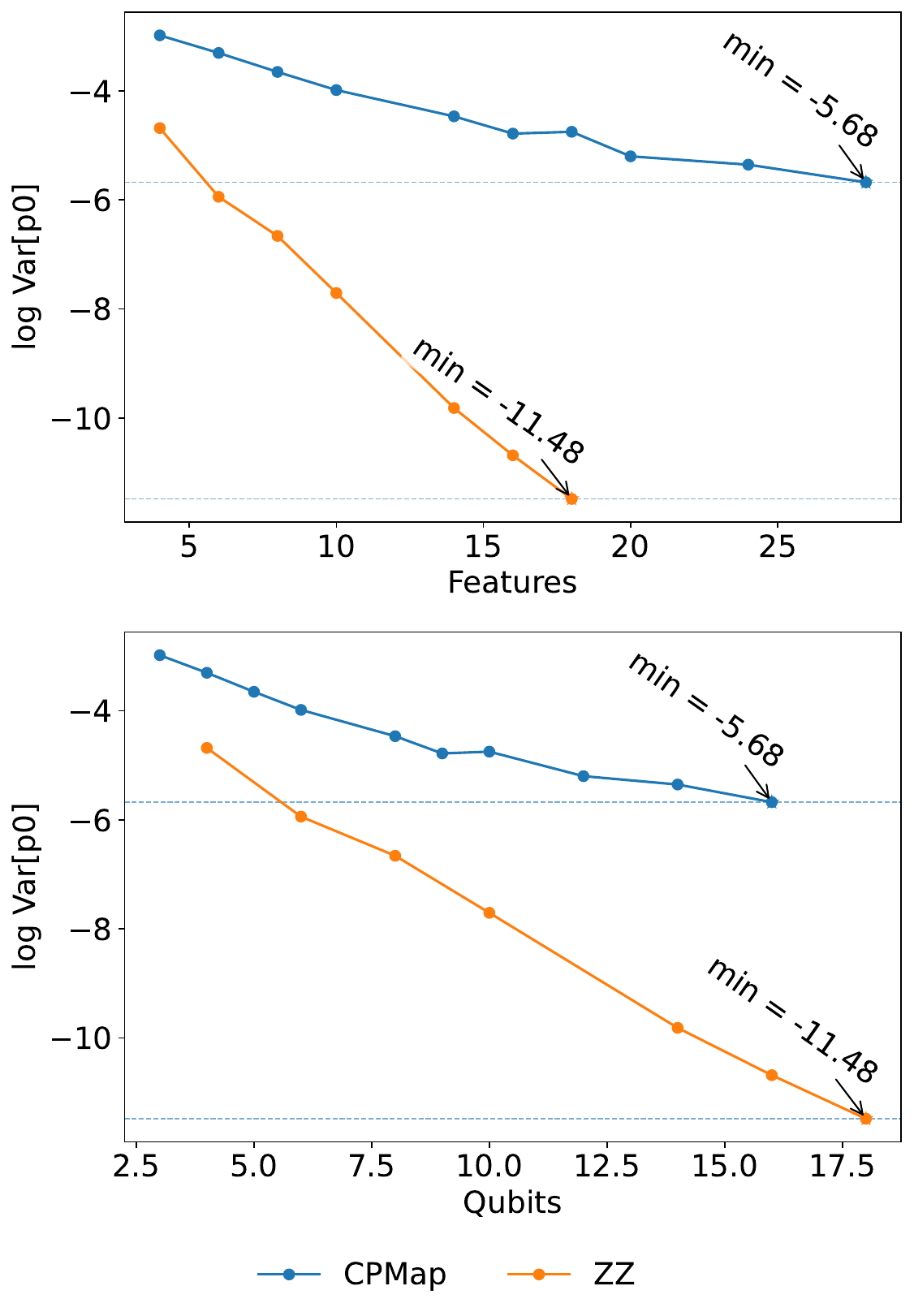}
    \caption{The plots compare the variance of kernel overlaps (shown as $\log \mathrm{Var}[p_0]$) across two feature maps: the proposed CPMap and the conventional ZZFeatureMap. (Top) Variance as a function of feature dimension $F$. (Bottom) Variance as a function of effective qubit count $n$. In both cases, the CPMap maintains consistently higher variance and delays the onset of exponential concentration relative to the ZZFeatureMap. These results show that the CPMap mitigates concentration effects and preserves discriminative power in higher-dimensional settings.}
    \label{fig:concentration}
\end{figure}

To analyze the issue of exponential concentration, we tested both feature maps (ZZFeatureMap and CPMap) with the fidelity quantum kernel on the \href{https://www.kaggle.com/datasets/jamieleech/ionosphere}{Ionosphere dataset}, a binary classification dataset commonly used in kernel-based learning. The diagnostic in Fig.~\ref{fig:concentration} (top) plots $\log \mathrm{Var}[p_{0}]$ as a function of feature dimension $F$, while the bottom panel shows the same diagnostic against the actual number of qubits used. At the same number of features, the ZZFeatureMap exhibits faster decay (minimum $\log \mathrm{Var}[p_{0}] \approx -11.5$) compared to CPMap (minimum $\approx -5.7$), indicating stronger concentration due to using more qubits per feature. Even when we compared with the same number of qubits, CPMap consistently maintains higher variance in overlap values than ZZ, implying that it retains discriminative power for a longer period and has a better decay rate $c$ by a factor of 2-3.

Additionally, from a practical point of view, extremely low variance (e.g., $\mathrm{Var}[p_0] \sim 10^{-11}$ for ZZFeatureMap) means we need an extremely large number of measurement shots (on the order of $10^{11}$) to have a meaningful discriminative power. This is not practical for the current devices. At the same time, CPMap has a consistently higher variance, which suggests it can obtain meaningful information with fewer shots, making it far more suitable for the NISQ era. 

Although CPMap does not eliminate the fundamental limitation of exponential concentration, it \emph{extends the useful operating regime} of fidelity kernels on near-term devices. By reducing the qubit and CNOT overhead, CPMap enables exploration of higher-dimensional datasets before entering the severe concentration regime, offering a practical pathway for applying quantum kernels in real-world machine learning settings.

\section{Discussion and Conclusion \label{secV}}

Our proposed feature map paves the way for the application of quantum machine learning algorithms to more complex and higher-dimensional data, even with limited qubit resources. The CPMap's remarkable efficiency in qubit utilization, evident through its requirement of approximately $\frac{N}{2}$ qubits to represent $N$ features, marks a substantial improvement over the ZZFeatureMap. This efficiency not only eases the computational load but also enhances the scalability of quantum models. In our comparative analysis with classical kernels such as the RBF, linear, and polynomial kernels, the CPMap sometimes demonstrated superior performance. Our findings show that, for the studied datasets and preprocessing choices, CPMap can achieve competitive (and sometimes higher) predictive performance than standard classical kernels, while using substantially fewer quantum resources than common Pauli feature maps.

The operational feasibility of the CPMap is another cornerstone of our research. Its reduced circuit depth, as opposed to the computationally intensive ZZFeatureMap, enabled us to successfully simulate complex datasets, due to the significant reduction in number of entangling gates required for our scheme. This aspect of the CPMap not only addresses the current limitations in quantum computing resources but also makes quantum machine learning models more accessible and practical for a broader range of applications.

\paragraph{Note on classical simulability.}
CPMap is intentionally shallow and locally structured to improve NISQ feasibility. The circuit structure underlying CPMap resembles QCNN-style architectures. Such architectural choices can also place a model in regimes where classical approximation methods remain effective. Recent work by Bermejo \cite{bermejo_simulable} argues that QCNNs can be \emph{effectively} classically simulable in practically relevant regimes, most notably, when their action is restricted to information accessible via low-bodyness observables and when the benchmark instances are ``locally-easy,'' so that a classical algorithm equipped with (Pauli) classical shadows can reproduce the relevant behaviour.

In contrast, our method uses CPMap as a \emph{feature map}, for kernel method, where training and evaluation require estimating state overlaps of the form $|\langle 0|U(y)^\dagger U(x)|0\rangle|^2$ across many pairs. Whether such overlap estimation is efficiently approximable classically can depend strongly on the circuit depth, entanglement growth, and the input distribution (e.g., the tensor-network bond dimension required for accurate contraction). We therefore do not make a general simulability or hardness claim here; rather, we position CPMap as a resource-efficient embedding and treat questions of classical simulability as regime-dependent.

The implications of this work extend beyond the specific benchmarks considered here. By reducing qubit requirements and limiting the growth of two-qubit entangling operations, CPMap makes kernel-based quantum learning experiments more accessible at higher feature counts, where many standard feature maps quickly become resource-intensive. This motivates several directions for future work, particularly in applying quantum algorithms to more complex and diverse datasets with applications in health sciences to material discovery. Further exploration into its realization on existing noisy quantum computing platforms and future small-scale fault-tolerant devices, as opposed to simulations, could offer deeper insights into its practical utility and performance in real-world applications.

Rather than positioning CPMap as a universal replacement for Pauli feature maps, we view it as a resource-efficient alternative that can be advantageous when qubit count and entangling-gate budgets are the dominant constraints. Our numerical benchmarks and small-scale hardware demonstrations indicate that CPMap can achieve competitive performance while using fewer qubits and fewer entangling gates (CNOTs) than standard Pauli-style constructions at comparable feature counts. These results suggest CPMap as a useful candidate for near-term kernel workflows and for systematic benchmarking against strong classical baselines. Clarifying the regimes where CPMap provides consistent benefits—across larger datasets, wider classes of quantum embeddings, and more realistic noise models—remains an important next step.

\section{Methods}

In this study, we introduce an instance of the CPMap, specifically designed for data classification using the kernel method across various standard datasets. The CPMap simulation is conducted utilizing two distinct environments: Qiskit's Statevector simulator and its Noisy counterpart, providing a comprehensive analysis of its performance under varied conditions. The datasets employed in this investigation are sourced from standard online repositories, namely Kaggle and the UCI dataset library, ensuring a diverse and robust set of data for evaluation. To facilitate the support vector classification process, we employ the built-in SVC function from Scikit-learn, a widely recognized tool in machine learning. This section details the parameters, methodologies, and evaluation techniques utilized in the development and assessment of the CPMap, aiming to demonstrate its effectiveness and versatility in data classification tasks.

To address the limitations of conventional metrics on imbalanced datasets (see Appendix~\ref{Appendix_A}), the MCC was employed. MCC is a balanced metric that takes into account both over-predictions and under-predictions across classes. It is defined as
\begin{equation}
\text{MCC} = \frac{TP \times TN - FP \times FN}{\sqrt{(TP + FP)(TP + FN)(TN + FP)(TN + FN)}},
\end{equation}
where TP, TN, FP, and FN are the numbers of true positives, true negatives, false positives, and false negatives, respectively. The MCC returns a value in $(-1,1)$, with 1 representing perfect prediction, -1 indicating total disagreement between prediction and observation, and 0 suggesting no better than random prediction.
MCC is particularly valuable for imbalanced datasets as it considers all four components of the confusion matrix and is less susceptible to the bias of a large class \cite{Chicco2017}. 

To facilitate a comprehensive comparison, we utilized seven distinct features for each dataset when evaluating the CPMap against the ZZFeatureMap. This choice is strategically made, considering that the ZZFeatureMap demands one qubit per feature and, notably, the number of CNOT gates required for the ZZFeatureMap increases quadratically with the number of features. Such rapid growth renders the simulation of larger datasets challenging. Opting for seven features strikes a balance, allowing for a meaningful comparison while keeping the computational requirements within manageable limits. In contrast to the ZZFeatureMap, the CPMap demonstrates remarkable efficiency in qubit utilization, requiring a mere four qubits to represent these seven features—a notable advancement over the ZZFeatureMap kernel. Moreover, we have benchmarked the CPMap's performance against several classical kernels, namely the RBF, linear, and polynomial kernels, to establish a baseline comparison.

For the noisy simulation, we targeted three distinct datasets: Stellar Classification, Balanced Credit Card, and Heart Disease. To accommodate computational limitations and manage extended runtimes, we applied PCA to reduce the dimensionality of the Heart Disease and Balanced Credit Card datasets, ultimately selecting the seven most informative features for these simulations. These noisy simulations were conducted using Qiskit's advanced noisy simulators, designed to mimic the behaviour of actual quantum computers, thereby providing a more realistic assessment of our models' performance in quantum computing environments. For the Balanced Credit Card and Heart Disease datasets, we utilized the \textbf{$\text{fake\_ibmq\_manila}$} backend, a simulator that emulates the noise characteristics of the IBMQ Manila quantum device. To avoid cherry picking, for the Stellar Classification dataset, we employed the \textbf{$\text{fake\_ibmq\_jakarta}$} backend, replicating the conditions of the IBMQ Jakarta, and for the Ionosphere dataset, we used \textbf{$\text{fake\_ibm\_perth}$}. These choices allowed us to assess the robustness of our models against realistic quantum noise and error rates. The results from these simulations, highlighting the impact of quantum noise on model accuracy and reliability, are detailed in Fig.~\ref{mcc_noisy}, providing critical insights into the potential real-world performance of quantum machine learning algorithms.

In our experiments, we also processed the datasets in their entirety using the CPMap without resorting to Principal Component Analysis (PCA) for dimensionality reduction. This approach provides a more authentic test of the CPMap's capability to handle high-dimensional data. Specifically, in the case of the Breast Cancer dataset, which requires sixteen qubits to represent thirty features, we observed that the CPMap necessitates significantly less circuit depth. This efficiency advantage enabled us to successfully simulate the model on a personal computer without encountering computational bottlenecks. It is crucial to note that running the Breast Cancer dataset with all 30 features using the ZZFeature Map was not feasible due to its computational expensiveness and the near impossibility of simulating 30 qubits with such high depth on personal computers. Notably, the quantum Kernel based on CPMap exhibited superior performance compared to the classical kernels in analyzing the Breast Cancer, Heart Disease datasets, thereby demonstrating its potential for practical applications in quantum machine learning and indicating a promising direction for further research in this area.

{We conducted experiments on IBM's superconducting quantum processors. Specifically, we tested our circuits on two real devices: ibm\_quebec (127-qubit Eagle-class) and ibm\_torino (133-qubit Heron-class). These evaluations were performed on the Parkinson’s disease dataset. On ibm\_quebec, we used all 22 features mapped to 12 qubits, as well as a PCA-reduced 7-feature input encoded on 4 qubits. For the 12 qubit case, error mitigation techniques including dynamical decoupling, zero noise extrapolation, and probabilistic error cancellation were applied in combination with varying transpilation levels to analyze hardware-induced deviations. On ibm\_torino, we performed an end-to-end kernel classification task using a curated subset of 100 samples selected via classical SVM pre-filtering, ensuring a mix of easy and ambiguous cases. We did not use any error mitigation technique for this task. The quantum kernel matrix was constructed in a single batch run, and training/testing splits were extracted without recomputation. These experiments allowed us to benchmark our kernel’s performance under real-device noise and confirm its competitiveness with classically optimized models, demonstrating the CPMap’s potential viability on current NISQ hardware.}

\section{Acknowledgments \label{secVI}}
The authors would like to acknowledge the use of IBM Quantum services for this work and in particular the Qiskit package \cite{Qiskit}, as well as fruitful discussions with Anaelle Hertz and Barry Sanders. We thank Marco Armenta for assisting with the execution of our experiments on the IBM Quebec quantum device. AZG and KH acknowledge that the NRC headquarters is located on the traditional unceded territory of the Algonquin Anishinaabe and Mohawk people. K.H. acknowledges funding
from the NSERC Discovery Grant and Alliance programs.

\section{Data Availability}
Data and code related to this research can be found at this private \href{https://github.com/utkarshh-singh/Quernel}{GitHub repository} upon reasonable request.

\newpage
\clearpage

\onecolumngrid
\begin{appendix}
    

\section{On the Suitability of Regular Machine Learning Metrics \label{Appendix_A}}

In the realm of machine learning, the evaluation of model performance is paramount. Common metrics such as accuracy, precision, recall, and F1 score are frequently employed to gauge the efficacy of models. However, these metrics, while widely accepted, are not universally applicable across all scenarios.

Imbalanced datasets, where one class significantly outnumbers the other, present a unique challenge for machine learning models and the evaluation metrics used to assess their performance. This is particularly true for binary classification tasks, where the minority class is often of greater interest than the majority class. 

\subsection{Limitations of Conventional Metrics}

Traditional performance metrics, such as accuracy, can be misleading in the context of imbalanced datasets \cite{Fawcett2006}. Consider a dataset with 95\% samples of class A and only 5\% samples of class B. A naive classifier predicting all samples as class A will still achieve a superficially high accuracy of 95\%. This demonstrates that accuracy alone is not sufficient to evaluate model performance on imbalanced datasets \cite{Chicco2020Dec,japkowicz2002}.

Similarly, other metrics such as recall, precision, and the F1 score can sometimes provide a skewed perspective when classes are imbalanced. There are situations where models can achieve high values for these metrics by being biased towards the majority class, rendering them less effective as measures of model performance \cite{Jeni2013,sokolova2009}.
This is why we employ the Matthews Correlation Coefficient (MCC), as described in the Methods section.

\subsection{Numerical Comparison:}

\begin{figure}[H]
    \centering
    \hspace*{-1cm}   
    \includegraphics[width=0.6\textwidth]{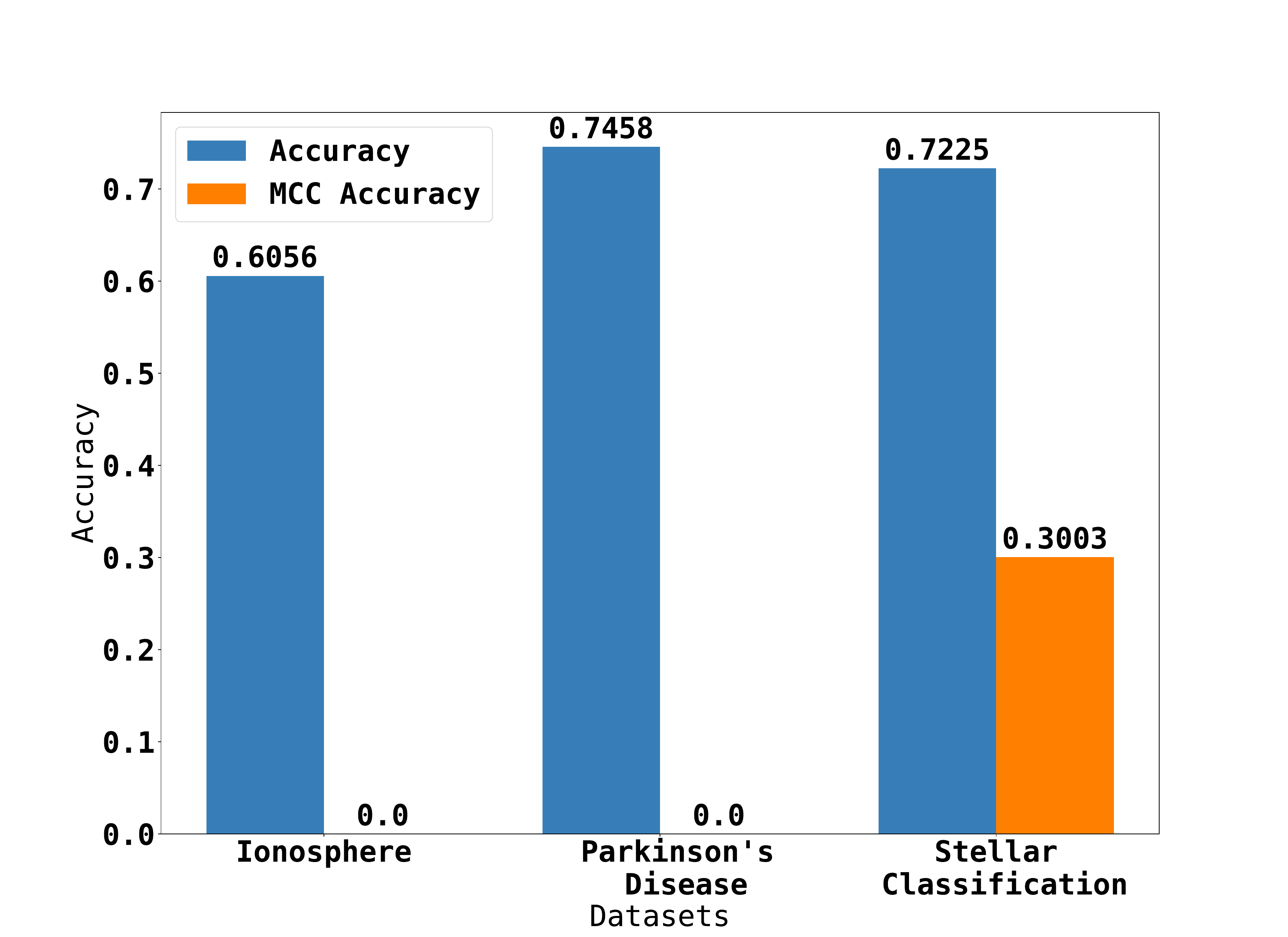}
    \caption{Evaluating accuracy (left bar, blue) vs. MCC (right bar, orange) across Ionosphere, Parkinson's Disease, and Stellar Classification datasets. Note that the MCC bar vanishes for the left two datasets.}
    \label{acc_vs_mcc}
\end{figure}

The bar plot (Fig.~\ref{acc_vs_mcc}) comparing the accuracy and MCC scores across the three datasets  Ionosphere, Parkinson's Disease, and Stellar Classification illustrates why the MCC score can be a more informative metric than accuracy, particularly in specific contexts such as imbalanced datasets or when true negatives are significant.

Firstly, in the Stellar Classification dataset, while the accuracy is relatively high (0.7225), the MCC score is notably lower (0.3003). This disparity suggests that although the model has a high rate of correct predictions (accuracy), its ability to balance true and false positives and negatives is less proficient, as captured by the MCC score. MCC takes into account all four categories of the confusion matrix (true positives, false positives, true negatives, and false negatives), offering a more balanced measure of the quality of binary classifications, especially in cases where class distribution is uneven.

Secondly, for the other two datasets, the MCC scores are zero despite the accuracies being 0.6056 and 0.7458, respectively. This indicates a situation where the model might be making correct predictions by chance or through bias towards the majority class, a common issue in imbalanced datasets. In such scenarios, accuracy alone can be misleading, as it does not distinguish between the types of errors made by the model. MCC, on the other hand, by being sensitive to the balance among all four confusion matrix categories, gives a more faithful representation of the model's performance.

\subsection{Conclusion}

When assessing machine learning models on datasets with imbalances, exclusive dependence on conventional metrics can result in skewed insights. The MCC presents a more encompassing and impartial assessment criterion, aptly fitting for tasks with class disparities. It is pivotal to select a performance measure that resonates with the distinct demands and traits of the problem at hand. Particularly in quantum machine learning, given the distinct nature of data and the nuances of the models, a mere reliance on metrics established for tasks other than the one at hand may fall short in truly capturing the model's efficacy.

\section{ZFeatureMap vs. ZZFeatureMap }
\label{Appendix_B}
The ZFeatureMap is a simple feature map that involves rotations around the $z$ axis of the Bloch sphere. On the other hand, the ZZFeatureMap is made from two-qubit gates that involve both rotations around the $z$ axis and controlled-X (CNOT) gates. This makes the ZZFeatureMap more expressive and capable of capturing more complex relationships in the data.

The primary reason for preferring the ZZFeatureMap over the ZFeatureMap in quantum machine learning is the former's ability to generate entanglement between qubits. Entanglement is a uniquely quantum phenomenon and is believed to be one of the reasons quantum algorithms can outperform classical ones \cite{Havlicek2019Mar,qiskit-textbook}. By using the ZZFeatureMap, quantum machine learning algorithms can leverage this entanglement to potentially achieve better performance.

While both the ZFeatureMap and ZZFeatureMap have their applications, the latter is generally preferred in quantum machine learning due to its ability to capture more complex data relationships and leverage quantum entanglement.

\section{Repeating the kernel to reupload the same data}
\subsection{Seven features}
\label{app:repetitions 7}
In Table~\ref{table3}, we present the efficacy of two quantum machine learning kernels, CPMap and ZZFeatureMap, across a variety of datasets, evaluated using the Matthews Correlation Coefficient (MCC). Each dataset was reduced to seven principal components, except for the Titanic dataset, where PCA was not applied, as indicated by an asterisk. The MCC scores serve as a quantitative measure of model performance, with a score of zero indicating no learning.

Regarding the number of repetitions (NOR), CPMap consistently outperformed ZZFeatureMap, even when only a single repetition was used. Specifically, the MCC scores for ZZFeatureMap dropped to zero for some datasets, suggesting no learning occurred, a situation not observed with CPMap. For ZZFeatureMap, the optimum performance was achieved with a single repetition, while additional repetitions led to a decrease in MCC scores, contrary to expectations. In contrast, CPMap's performance was enhanced with repetitions, as demonstrated by higher MCC scores across several datasets. This pattern also underscores the variable impact of the number of repetitions on different algorithms and the importance of tailored approaches in model training.

\begin{table*}[h!]
\centering
\begin{tabular}{|l|l|l|l|l|l|l|}
\hline
\textbf{Dataset} & \textbf{NOF Used} &  \textbf{NOR  for} & \textbf{NOR  for} & \textbf{MCC Score} & \textbf{MCC Score} & \textbf{MCC Score}\\
 &  &  \textbf{CPMap } & \textbf{ZZFeatureMap} & \textbf{CPMap } & \textbf{ZZFeatureMap} &  \textbf{CPMap with 1 Reps} \\
\hline
\href{https://www.kaggle.com/datasets/jamieleech/ionosphere}{Ionosphere} & 7 & 1 & 1 & 0.643 &  0.0 & 0.643\\
\hline
\href{https://archive.ics.uci.edu/dataset/17/breast+cancer+wisconsin+diagnostic}{Breast cancer Diagnostic} & 7 & 2 & 1 & 0.944 & 0.869 & 0.926\\
\hline
\href{https://www.kaggle.com/datasets/mlg-ulb/creditcardfraud}{Credit card (Balanced)} & 7 & 1& 1 & 0.831 & 0.602 & 0.831 \\
\hline
\href{https://archive.ics.uci.edu/dataset/174/parkinsons}{Parkinson's disease (PD)} & 7 & 2 & 1 & 0.520 & 0.0 & 0.451\\
\hline
\href{https://www.kaggle.com/datasets/fedesoriano/stellar-classification-dataset-sdss17}{Stellar Classification} & 7 & 2 & 1 & 0.791 & 0.300 & 0.778 \\
\hline
\href{http://archive.ics.uci.edu/dataset/45/heart+disease}{Heart Disease} & 7 & 2 & 1 & 0.664 & 0.449 & 0.621 \\
\hline
\href{https://www.kaggle.com/datasets/yasserh/titanic-dataset}{Titanic} & $7^{*}$ & 1 & 1 & 0.561 & 0.351 & 0.561\\
\hline
\end{tabular}
\caption{Data and model characteristics for datasets with seven features. NOF: Number of features; NOR: Number of repetitions; {*}without PCA.}
\label{table3}
\end{table*}

\subsection{All features}
\label{app:repetitions full}
In Table \ref{table4}, we present the efficacy of the CPMap in learning tasks across six datasets, emphasizing the model's robustness without substantial feature reduction, each with varying requirements for quantum resources, especially the number of repetitions of the CPMap circuits. The datasets under consideration included Ionosphere, Breast Cancer Diagnostic, Credit Card (Balanced), Parkinson's Disease (PD), Stellar Classification, and Agaricus Lepiota Mushrooms. The number of features utilized ranged from 16 to 30, with an asterisk indicating datasets where Principal Component Analysis (PCA) was not employed. The quantum resource requirement, measured in qubits, was predominantly 12, except for Stellar Classification, which required 9. The assessment metric, MCC, was reported for multiple repetitions of the CPMap and a singular repetition to investigate the impact of repetition on performance. Notably, the MCC scores exhibited a broad spectrum, with Parkinson's Disease recording the lowest at 0.672, while Agaricus Lepiota Mushrooms achieved an impressive high of 0.996. This variance underscores the model's performance sensitivity to the inherent characteristics of the dataset. 

\begin{table*}[h]
\centering
\begin{tabular}{|l|l|l|l|l|l|l|}
\hline
\textbf{Dataset} & \textbf{NOF Used} &  \textbf{NOR  for} & \textbf{Qubits} &   \textbf{MCC Score} & \textbf{MCC Score} & \textbf{Accuracy} \\
 &  &  \textbf{CPMap} & \textbf{Required} & & \textbf{with 1 repetition} & \\
\hline
\href{https://www.kaggle.com/datasets/jamieleech/ionosphere}{Ionosphere} & 22 & 2 & 12 & 0.866 & 0.832 & 0.930 \\
\hline
\href{https://archive.ics.uci.edu/dataset/17/breast+cancer+wisconsin+diagnostic}{Breast cancer Diagnostic} & $30^{*}$ & 2 & 16 & 0.943 & 0.908 & 0.974 \\
\hline
\href{https://www.kaggle.com/datasets/mlg-ulb/creditcardfraud}{Credit card
} (Balanced) & 22 & 5 & 12 & 0.950 & 0.942 & 0.934 \\
\hline
\href{https://archive.ics.uci.edu/dataset/174/parkinsons}{Parkinson's disease (PD)} & $22^{*}$ & 1 & 12 & 0.672 & 0.672 & 0.881 \\
\hline
\href{https://www.kaggle.com/datasets/fedesoriano/stellar-classification-dataset-sdss17}{Stellar Classification} & 16 & 3 & 9 & 0.956 & 0.946 & 0.982 \\
\hline
\href{https://www.kaggle.com/datasets/uciml/mushroom-classification}{Agaricus Lepiota Mushrooms} & $21^{*}$ & 6 & 12 & 0.996 & 0.910 & 0.998 \\

\hline
\end{tabular}
\caption{Data and model characteristics for datasets with all features. NOF: Number of features; NOR: Number of repetitions; {*}without PCA.}
\label{table4}
\end{table*}

\section{Results for higher number of features }
\label{app:full}
This section presents findings using exclusively the CPMap and a greater number of features, including the full set. In this comparison, the ZZFeatureMap was not utilized, as simulation attempts on our computers consistently resulted in memory errors. This suggests that employing ZZFeatureMap is computationally demanding.

\begin{figure}[H]
    \centering
    \hspace*{-1cm}   
    \includegraphics[width=0.9\textwidth]{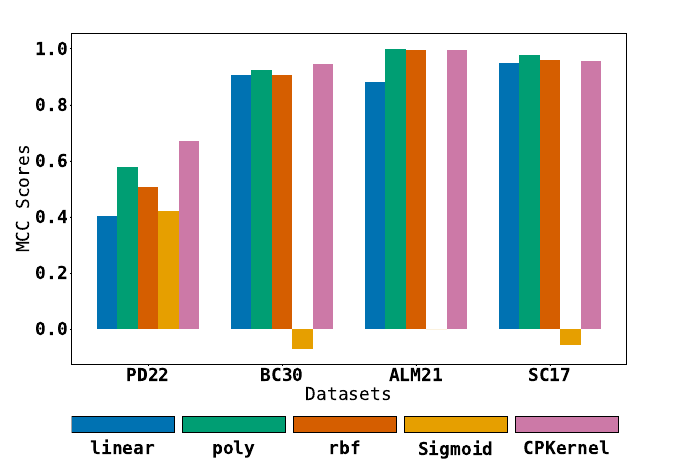}
    \caption{Comparative Analysis of MCC Scores for various datasets with all the Features, including Parkinson's Disease with 22 features (PD22), Breast Cancer with 30 features (BC30), Agaricus Lepiota Mushroom with 21 features (ALM21) and Stellar Classification with 17 features (SC17).}
    \label{allfeat}
\end{figure}

Figure \ref{allfeat} presents a comparison of MCC scores for several datasets, with all the features, encompassing Parkinson’s Disease with 22 features (PD22), Breast Cancer with 30 features (BC30), Agaricus Lepiota Mushroom with 21 features (ALM21), and Stellar Classification with 17 features (SC17). For the Stellar Classification dataset specifically, the analysis was focused on two classes, totaling 81,039 entries. The bar graph demonstrates that CPMap consistently achieves high MCC scores across these datasets, showcasing its capacity to effectively manage and interpret datasets with a substantial number of features. These results indicate that CPMap has the potential to surpass traditional kernel functions, affirming its suitability for complex, large-scale data analysis.

\section{Comparison with Data Re-Uploading Feature Maps}
\label{data_reuploading_fm}

In the main text, we benchmark CPMap against the widely used ZZFeatureMap~\cite{Havlicek2019Mar}. To further strengthen the positioning of CPMap within the broader landscape of resource-efficient quantum embeddings,  we additionally benchmark CPMap against six fixed (non-trained) circuit variants inspired by the data re-uploading framework of P\'erez-Salinas \emph{et al.}~\cite{PerezSalinas2020datareuploading}. Data re-uploading was originally introduced as a \emph{trainable} model in which classical inputs are injected at multiple layers and trainable parameters are optimized. Here, to remain within the \emph{fixed-embedding} kernel setting studied throughout this paper, we use the same circuit templates but \emph{freeze} the additional rotation offsets: the offsets are sampled once at initialization and then held fixed for all runs. 
This yields a deterministic kernel matrix for a given dataset and initialization, enabling a controlled comparison of encoding strategies without introducing additional trainable degrees of freedom.

\begin{figure}[!t]
    \centering
    \hspace*{-1cm}
    \includegraphics[width=0.9\textwidth]{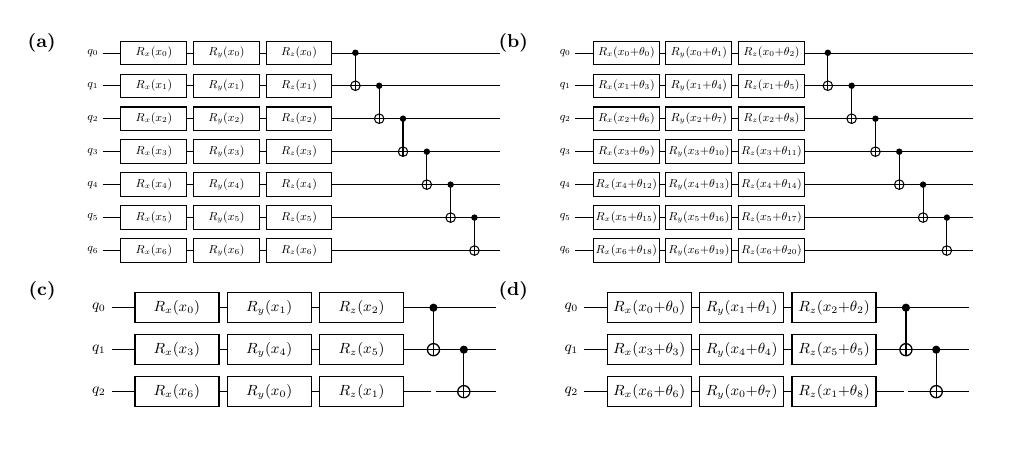}
    \caption{%
        Base circuit blocks used to construct the frozen data re-uploading
        feature maps DR1--DR6.
        \textbf{(a)}~Circuit~1 (7 qubits): one-qubit-per-feature encoding
        with successive single-qubit rotations per feature and a linear
        entangling chain.
        \textbf{(b)}~Circuit~1 with frozen random offsets $\boldsymbol{\theta}$
        added to rotation angles (random-feature style kernel).
        \textbf{(c)}~Circuit~2 (3 qubits): feature packing that distributes
        the $d=7$ features across qubit--rotation pairs via modular indexing,
        followed by linear entanglement.
        \textbf{(d)}~Circuit~2 with frozen random offsets.
        DR1--DR6 are obtained by repeating these base blocks three or six
        times as described in the text.%
    }
    \label{fig:data_reupload_circuits}
\end{figure}

\textbf{Base circuits and DR variants:}
We consider two base encoding blocks (Fig.~\ref{fig:data_reupload_circuits}) designed to embed a $d=7$ feature vector under different resource constraints. \textbf{Circuit~1} uses a one-qubit-per-feature strategy across 7 qubits: each qubit $q_i$ encodes feature $x_i$ via a sequence of single-qubit rotations (e.g., $R_x, R_y, R_z$), followed by a linear entangling chain to introduce inter-qubit correlations. \textbf{Circuit~2} implements a feature-packing strategy, encoding all 7 features into only 3 qubits by distributing feature indices across qubit-rotation pairs using modular indexing, again followed by a linear entangling pattern. For each base circuit we consider two parameterizations: a \emph{parameter-free} version and a \emph{frozen-offset} version in which rotation angles are augmented by fixed random offsets $\boldsymbol{\theta}$ sampled once and then held constant.

From these blocks we define six re-uploading feature maps DR1--DR6 by repeating the base blocks multiple times: DR1/DR2 repeat Circuit~1 three and six times, respectively; DR3/DR4 repeat the frozen-offset variant of Circuit~1 three and six times; DR5/DR6 repeat Circuit~2 three times (with and without frozen offsets). The depth-matched variants DR2 and DR4 were included specifically to match CPMap's circuit depth (34 in the $d=7$ setting), so that performance differences can be interpreted as differences in encoding structure rather than simply additional circuit depth. Circuit resources for all feature maps are summarized in Table~\ref{tab:dr_circuit_resources}. In all cases, the quantum kernel is given by \ref{eqn:11}, and is used as a precomputed kernel in a support vector classifier; all learning takes place in the classical SVC on top of the fixed quantum feature map.
\begin{table}[H]
\centering
\caption{Quantum circuit resources required by CPMap and the six frozen data re-uploading feature maps (DR1--DR6) for a $d=7$ input. Circuit depth and CX gate count are reported for the full feature map after all repetitions. DR2 and DR4 are depth-matched to CPMap (depth 34).}
\label{tab:dr_circuit_resources}
\begin{tabular}{lcccc}
\toprule
\textbf{Feature Map} & \textbf{Qubits} & \textbf{Depth} & \textbf{CX Count} & \textbf{Repetitions} \\
\midrule
CPMap  & 7 & 34 & 21 & 1 \\
\midrule
DR1    & 7 & 19 & 18 & 3 \\
DR2    & 7 & 34 & 36 & 6 \\
DR3    & 7 & 19 & 18 & 3 \\
DR4    & 7 & 34 & 36 & 6 \\
\midrule
DR5    & 3 & 15 &  6 & 3 \\
DR6    & 3 & 15 &  6 & 3 \\
\bottomrule
\end{tabular}
\end{table}

\subsection*{Classification results}
\begin{figure}[H]
    \centering
    \includegraphics[width=0.9\linewidth]{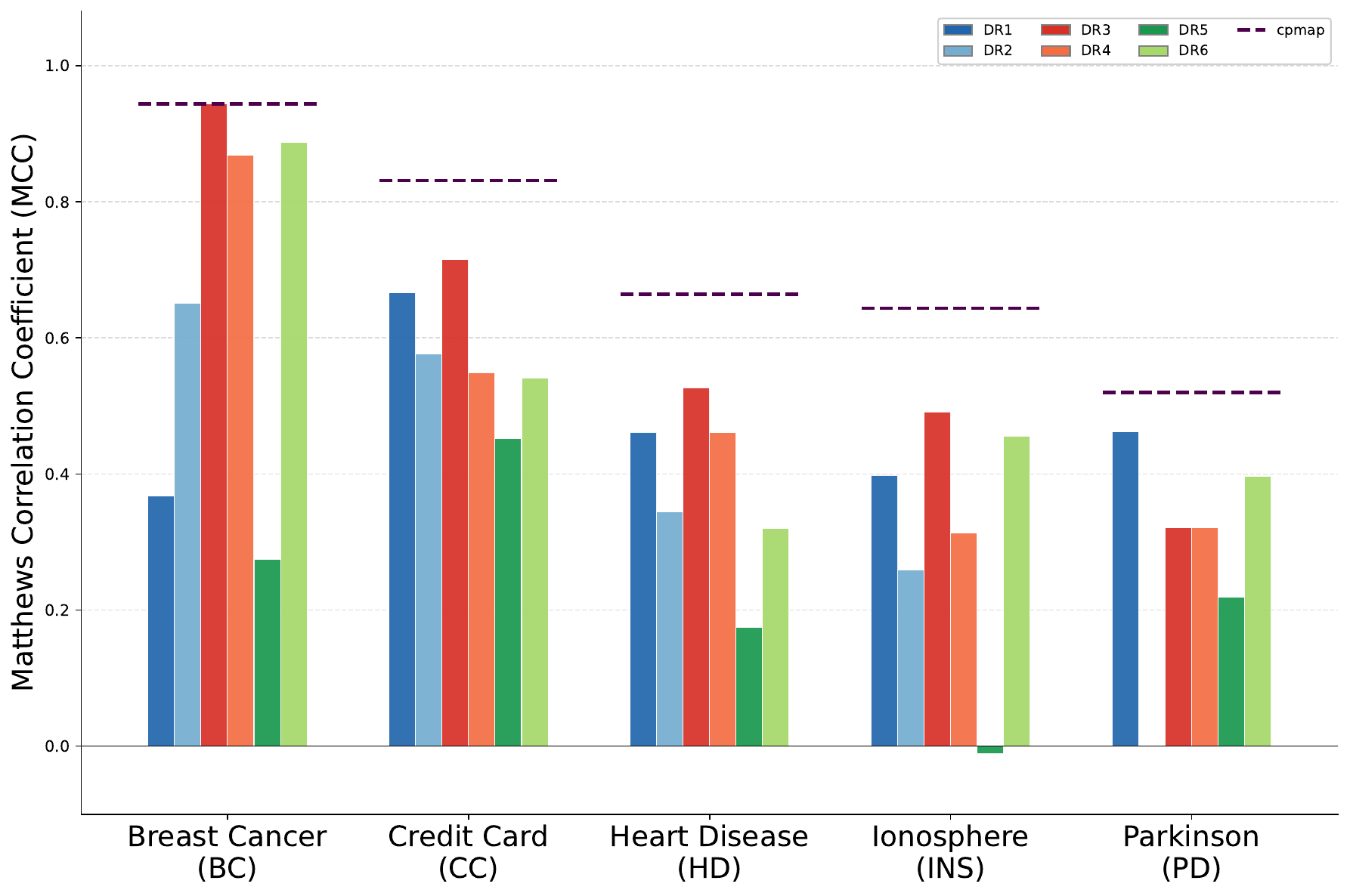}
    \caption{%
        MCC achieved by CPMap (dashed line) and the frozen data re-uploading feature maps DR1--DR6 across five benchmark datasets. Each configuration is evaluated on the same fixed train-test split. Higher MCC indicates stronger classification performance; values near zero correspond to near-random predictions.%
    }
    \label{fig:dr_mcc}
\end{figure}

We evaluate CPMap and the six DR variants on five benchmark datasets
using the Matthews correlation coefficient (MCC), which is robust to class imbalance and summarizes all entries of the confusion matrix.
Figure~\ref{fig:dr_mcc} reports the resulting MCC values.

Across the evaluated datasets, CPMap matches or exceeds the performance of all six frozen re-uploading variants. On \textbf{Breast Cancer}, the strongest DR variant reaches MCC $=0.944$, matching CPMap; this is the only dataset where a generic re-uploading construction fully recovers CPMap's performance. On \textbf{Credit Card}, \textbf{Heart Disease}, \textbf{Ionosphere}, and \textbf{Parkinson's Disease}, CPMap exceeds the best DR variant by $0.165$, $0.137$, $0.245$, and $0.124$ MCC points, respectively. Notably, the depth-matched variants DR2 and DR4 (depth 34) do not consistently improve over their shallower counterparts DR1 and DR3, and remain below CPMap on all datasets except Breast Cancer. This observation is important: it indicates that the performance gap is not explained merely by additional depth in a re-uploading architecture, but reflects the benefit of CPMap's structured encoding strategy under resource constraints.

\paragraph{Takeaway:}
These additional baselines strengthen the main conclusion of the paper: CPMap provides a competitive, resource-efficient kernel embedding that remains robust when compared not only to standard Pauli-style feature maps, but also to deeper re-uploading-style embeddings in a controlled (fixed, non-trained) setting.

\section{Multi-class evaluation on the Stellar Classification dataset \label{multiclass}}

To assess whether CPMap remains effective beyond binary classification, we additionally evaluate it on a genuine \emph{three-class} task using the Stellar Classification dataset. We construct a balanced subset by sampling 500 examples from each class (total $N=1500$) and follow the same preprocessing protocol used elsewhere in the paper: features are standardized and then reduced to $d=12$ components using PCA fit on the training and test splits. We then compute the CPMap kernel matrices using the same fixed CPMap hyperparameters as in the main experiments and train a multi-class SVM with a precomputed kernel using a one-vs-rest strategy:
\[
\texttt{SVC(kernel="precomputed", decision\_function\_shape="ovr", C=C\_SVM)}.
\]
On this 3-class evaluation, CPMap achieves the following performance:
\[
\text{MCC}=0.650,\qquad \text{Balanced Accuracy}=0.767,\qquad
F1_{\text{macro}}=0.768,\qquad F1_{\text{weighted}}=0.768.
\]
These results provide evidence that CPMap can be applied to multi-class settings without modifying the feature map design or its hyperparameters, complementing the binary benchmarks reported in the main text.

\section{CPMap hyperparameters and robustness \label{hyperparam}}

\subsection*{CPMap hyperparameters and default setting}
CPMap is specified by a small set of fixed design angles that parameterize the two-qubit interaction blocks used in the feature map. These angles are not trained and therefore function as hyperparameters. We denote the full CPMap hyperparameter vector by
\begin{equation}
\Theta_{\mathrm{CP}} = (\alpha_C,\beta_C,\gamma_C,\alpha_P,\beta_P,\gamma_P),
\end{equation}
where $(\alpha_C,\beta_C,\gamma_C)$ parameterize the $C$-block and $(\alpha_P,\beta_P,\gamma_P)$ parameterize the $P$-block (see main text for the circuit definition).

Unless stated otherwise, we use the default setting
\begin{equation}
(\alpha_C,\beta_C,\gamma_C,\alpha_P,\beta_P,\gamma_P)
=\left(-\frac{\pi}{3},\frac{\pi}{6},-\frac{\pi}{9},\frac{\pi}{7},\frac{\pi}{9},-\frac{\pi}{7}\right),
\end{equation}
which is a fixed, non-trained configuration chosen prior to benchmarking. The purpose of this section is to verify that CPMap does not rely on a finely tuned hyperparameter choice and that parameter variations induce systematic changes in the induced kernel geometry.

All hyperparameter analyses use the Breast Cancer dataset restricted to a fixed $n=100$ subset (same subset across all experiments) and the same kernel SVM evaluation pipeline. For each hyperparameter setting we evaluate performance over $50$ stratified train/test splits and report the mean Matthews correlation coefficient (MCC) with $95\%$ confidence intervals (CI). Unless stated otherwise, all other aspects of the pipeline (preprocessing, feature dimension, and circuit depth) are held fixed to isolate the effect of the CPMap angles.

\subsubsection{Why we set local KAK factors to the identity.}
In the Cartan/KAK form $U=(u_1\!\otimes\!u_2)\exp[-\frac{i}{2}(c_x X\!\otimes\!X+c_y Y\!\otimes\!Y+c_z Z\!\otimes\!Z)](v_1\!\otimes\!v_2)$, the local unitaries $(u_1,u_2,v_1,v_2)$ do not affect the nonlocal invariants $(c_x,c_y,c_z)$ but they add additional single-qubit layers. In CPMap we set these local factors to the identity to obtain a canonical minimal representative of each interaction family (isotropic/XY/ZZ) and to keep the feature map as resource-efficient as possible. This choice reduces depth and calibration overhead and keeps the embedding in the ``fixed, non-trained'' regime considered in this work. The sensitivity studies in Supplementary A-B show that CPMap remains robust and that its remaining angles systematically modulate kernel geometry and performance.

\subsection*{Structured parameter sweeps: sensitivity of performance to block parameters}
To test robustness in a controlled manner, we perform structured one-parameter sweeps over a representative sweep angle $\theta$ within each of three interaction families (isotropic, XY-type, and ZZ-type). We examine sensitivity separately for the two CPMap blocks:
\begin{itemize}
    \item \textbf{Varying the C-block:} we sweep the designated C-block angle while holding the remaining angles (including all P-block angles) fixed.
    \item \textbf{Varying the P-block:} we sweep the designated P-block angle while holding the remaining angles (including all C-block angles) fixed.
\end{itemize}

\begin{tcolorbox}[
  enhanced,
  colback=gray!4,
  colframe=black!60,
  boxrule=0.6pt,
  arc=2mm,
  left=2mm,right=2mm,top=1.5mm,bottom=1.5mm
]
\paragraph{Interaction-family terminology (isotropic, XY-type, ZZ-type).}
In our structured sweep we group two-qubit blocks into three \emph{interaction families}, which describe the effective Pauli-coupling content of the entangling operation. Concretely, any two-qubit unitary can be expressed (up to local unitaries) via a Cartan/KAK form
\begin{equation}
U \;\equiv\; (u_1 \otimes u_2)\,
\exp\!\left[-\frac{i}{2}\left(
c_x\, X\!\otimes\!X \;+\; c_y\, Y\!\otimes\!Y \;+\; c_z\, Z\!\otimes\!Z
\right)\right]\,
(v_1 \otimes v_2),
\label{eq:kak}
\end{equation}
where $X,Y,Z$ are Pauli operators and $(u_1,u_2,v_1,v_2)$ are single-qubit unitaries. The triple $(c_x,c_y,c_z)$ determines the nonlocal ``interaction type'' of the gate.

\begin{itemize}
    \item \textbf{Isotropic (balanced mixing).} A balanced nonlocal coupling with comparable strengths along the three Pauli axes, i.e., $c_x \approx c_y \approx c_z$.
    \item \textbf{XY-type.} Exchange-like coupling dominated by $X\!\otimes\!X$ and $Y\!\otimes\!Y$, i.e., $c_x \approx c_y \neq 0$ and $c_z \approx 0$.
    \item \textbf{ZZ-type.} Ising-like coupling dominated by $Z\!\otimes\!Z$, i.e., $c_z \neq 0$ with $c_x \approx c_y \approx 0$.
\end{itemize}

\noindent
In the sweep experiments, ``isotropic/XY/ZZ'' indicates which Pauli-coupling content we vary within the two-qubit block, while keeping the remainder of the feature map fixed. This terminology is used only to concisely describe the dominant nonlocal generator of the chosen two-qubit sub-block and does not imply that the physical device natively implements these Hamiltonians.
\end{tcolorbox}

\begin{figure}[!t]
    \centering
    \includegraphics[width=\linewidth]{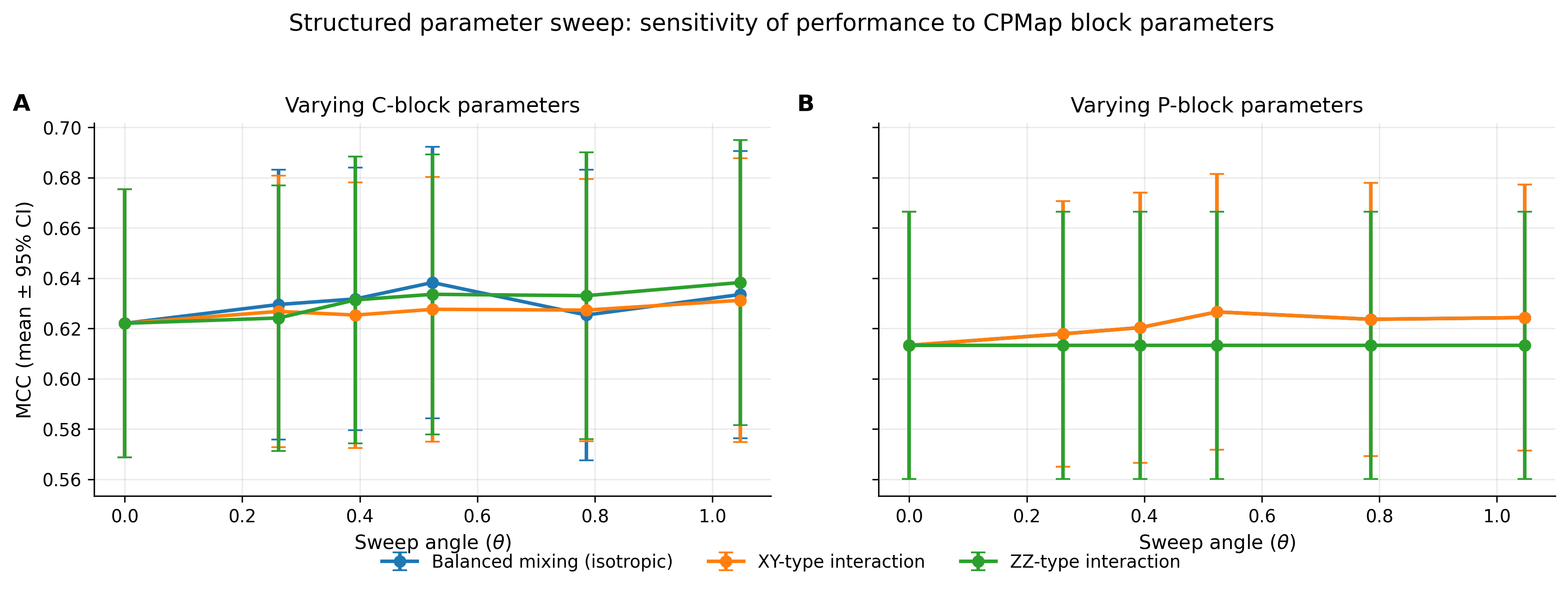}
    \caption{\textbf{Structured parameter sweep: sensitivity of performance to CPMap block parameters (Breast Cancer, $n=100$).}
    MCC (mean $\pm$ 95\% CI over 50 stratified splits) as a function of sweep angle $\theta$ for three interaction families. 
    (A) Varying the C-block while keeping the P-block fixed. (B) Varying the P-block while keeping the C-block fixed.
    The relatively flat response across a wide range of $\theta$ indicates robustness to hyperparameter variation.}
    \label{fig:S1_structured_sweep}
\end{figure}

Figure \ref{fig:S1_structured_sweep} reports MCC (mean $\pm$ 95\% CI) as a function of $\theta$ for each interaction family. Across both blocks and all families, performance varies smoothly and remains within a narrow band over a broad range of $\theta$. This indicates that CPMap does not depend on a single finely tuned parameter value and that moderate deviations from the default configuration do not cause performance collapse.

\subsection*{Paired comparison: improvements are consistent across splits}
\begin{figure}[H]
    \centering
    \includegraphics[width=0.8\linewidth]{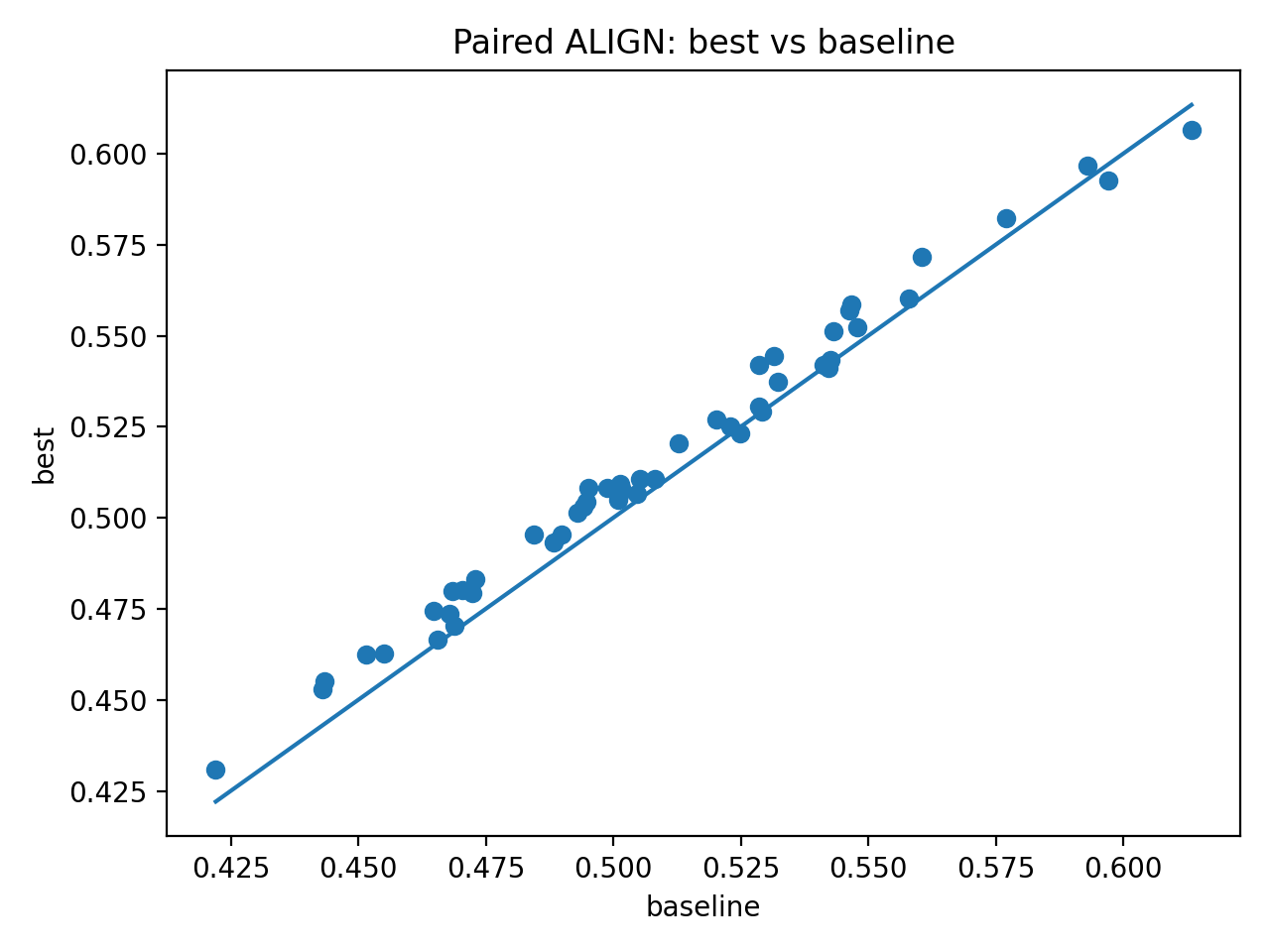}
    \caption{\textbf{Paired alignment comparison: best vs baseline (Breast Cancer, $n=100$).}
    Each point corresponds to one of the 50 stratified splits, showing the centered kernel--label alignment of a selected strong hyperparameter configuration (y-axis) versus the baseline configuration (x-axis). Points above the diagonal indicate splits where the selected configuration yields higher alignment. The consistent upward shift indicates robust improvement across splits.}
    \label{fig:S3_paired_align}
\end{figure}

To verify that improvements are not driven by a small number of favorable splits, we perform a paired split-wise comparison of alignment between a selected strong CPMap configuration (\emph{best}) and the default configuration (\emph{baseline}) using the same 50 splits. Figure~\ref{fig:S3_paired_align} plots split-wise alignment values (best vs baseline). Most points lie above the diagonal, indicating that alignment improvements are consistent across splits rather than arising from outliers. This supports the conclusion that CPMap hyperparameters control kernel geometry in a reproducible way.

\noindent\textbf{Summary.} The structured sweeps (Fig.~S1) demonstrate that CPMap is not brittle with respect to its fixed design angles. The random screening (Fig.~S2) shows that these angles systematically modulate kernel geometry and performance, and the paired analysis (Fig.~S3) confirms that the effect is consistent across splits. Together these results address the concern that CPMap hyperparameters are ad-hoc by establishing robustness and providing an interpretable diagnostic (alignment) for selecting reasonable parameter regimes.

\section{Coherence feasibility and runtime interpretation (hardware)}
\label{app:coherence_runtime}

\begin{figure}[H]
    \centering
    \includegraphics[width=\linewidth]{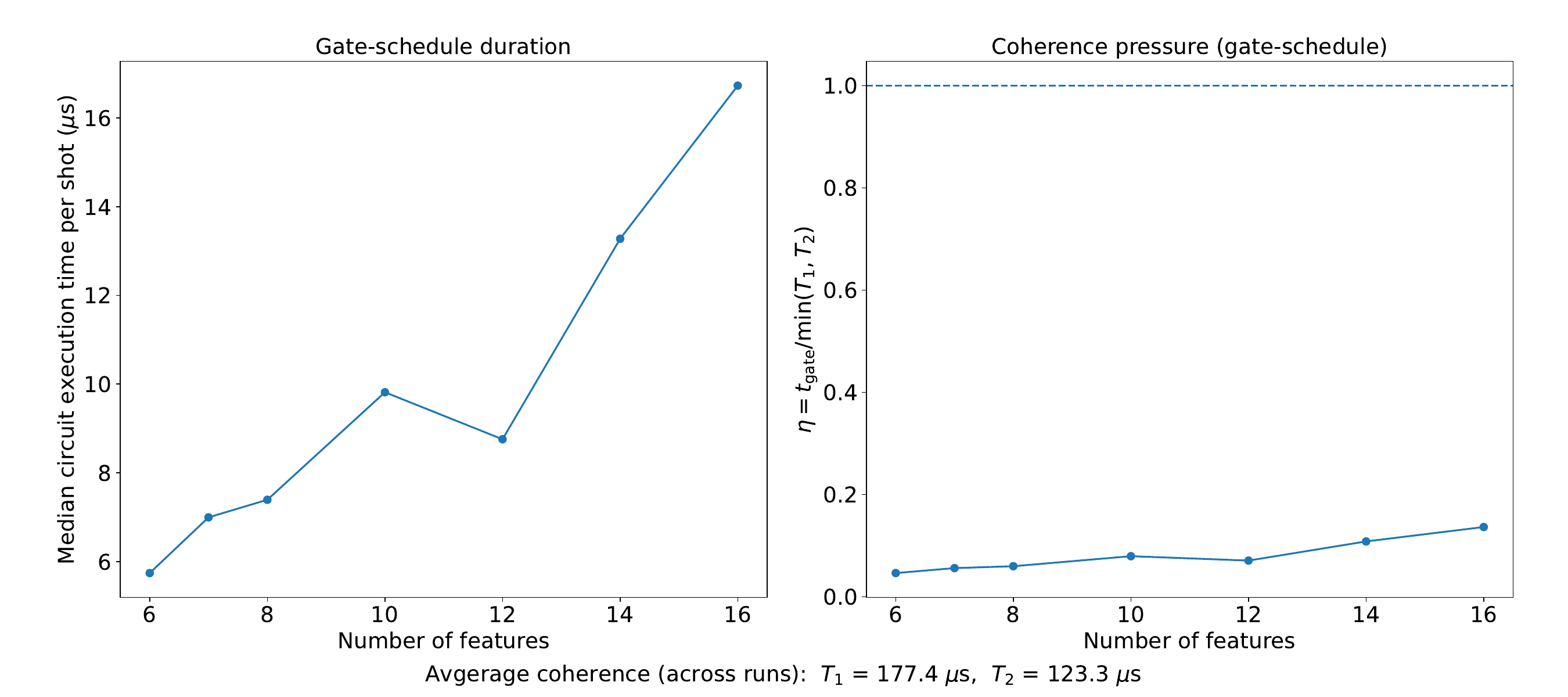}
    \caption{\textbf{Gate-schedule duration and coherence pressure for hardware kernel circuits.}
    \textbf{(Left)} Median estimated gate-schedule duration per shot (in $\mu$s) of the transpiled circuits as a function of the number of encoded features. Each point summarizes the median across the batch of circuits executed for that feature dimension.
    \textbf{(Right)} Corresponding coherence-pressure ratio $\eta = t_{\mathrm{gate}}/\min(T_1,T_2)$, where $t_{\mathrm{gate}}$ is the median gate-schedule duration per shot and $T_1$, $T_2$ are the backend relaxation and dephasing times (median across runs; values reported below the panels). The dashed line at $\eta=1$ marks the boundary where the scheduled circuit duration matches the coherence window. Across all tested feature dimensions, $\eta \ll 1$, indicating that the compiled schedules remain well within the coherence time budget; performance degradation at larger feature dimensions is therefore unlikely to be explained by circuit duration exceeding coherence.}
    \label{fig:coherence}
\end{figure}

\paragraph{Gate-schedule duration vs.\ coherence.}
To assess whether the executed circuits are feasible within the coherence window of current superconducting hardware, we use the backend timing model to estimate the \emph{gate-schedule duration} per shot for each compiled circuit. Concretely, after transpiling each overlap circuit to the target backend, we extract the estimated duration using Qiskit's timing estimator (reported in our logs as \texttt{duration\_median\_s}) and summarize it by the median across the 100 circuits in a batch. Figure~\ref{fig:coherence} (left) reports the resulting median per-shot circuit execution time (in $\mu$s) as a function of feature dimension. We compare this time scale to the device coherence times using the ratio
\begin{equation}
\eta \;=\; \frac{t_{\mathrm{gate}}}{\min(T_1, T_2)},
\end{equation}
where $t_{\mathrm{gate}}$ denotes the median gate-schedule duration per shot and $T_1$, $T_2$ are the median relaxation and dephasing times from the backend properties. Figure~\ref{fig:coherence} (right) shows that $\eta$ remains well below unity across all tested dimensions (dashed line at $\eta=1$), indicating that the compiled gate schedule fits comfortably within the coherence window. Therefore, the degraded performance observed at larger feature dimensions is unlikely to be explained by circuit duration exceeding coherence, and is more consistent with error accumulation from two-qubit gates, readout error, crosstalk, and other gate-level noise mechanisms.

\end{appendix}


\begin{thebibliography}{35}%
\makeatletter
\providecommand \@ifxundefined [1]{%
 \@ifx{#1\undefined}
}%
\providecommand \@ifnum [1]{%
 \ifnum #1\expandafter \@firstoftwo
 \else \expandafter \@secondoftwo
 \fi
}%
\providecommand \@ifx [1]{%
 \ifx #1\expandafter \@firstoftwo
 \else \expandafter \@secondoftwo
 \fi
}%
\providecommand \natexlab [1]{#1}%
\providecommand \enquote  [1]{``#1''}%
\providecommand \bibnamefont  [1]{#1}%
\providecommand \bibfnamefont [1]{#1}%
\providecommand \citenamefont [1]{#1}%
\providecommand \href@noop [0]{\@secondoftwo}%
\providecommand \href [0]{\begingroup \@sanitize@url \@href}%
\providecommand \@href[1]{\@@startlink{#1}\@@href}%
\providecommand \@@href[1]{\endgroup#1\@@endlink}%
\providecommand \@sanitize@url [0]{\catcode `\\12\catcode `\$12\catcode `\&12\catcode `\#12\catcode `\^12\catcode `\_12\catcode `\%12\relax}%
\providecommand \@@startlink[1]{}%
\providecommand \@@endlink[0]{}%
\providecommand \url  [0]{\begingroup\@sanitize@url \@url }%
\providecommand \@url [1]{\endgroup\@href {#1}{\urlprefix }}%
\providecommand \urlprefix  [0]{URL }%
\providecommand \Eprint [0]{\href }%
\providecommand \doibase [0]{https://doi.org/}%
\providecommand \selectlanguage [0]{\@gobble}%
\providecommand \bibinfo  [0]{\@secondoftwo}%
\providecommand \bibfield  [0]{\@secondoftwo}%
\providecommand \translation [1]{[#1]}%
\providecommand \BibitemOpen [0]{}%
\providecommand \bibitemStop [0]{}%
\providecommand \bibitemNoStop [0]{.\EOS\space}%
\providecommand \EOS [0]{\spacefactor3000\relax}%
\providecommand \BibitemShut  [1]{\csname bibitem#1\endcsname}%
\let\auto@bib@innerbib\@empty
\bibitem [{\citenamefont {Chen}(2009)}]{Chen2009}%
  \BibitemOpen
  \bibfield  {author} {\bibinfo {author} {\bibfnamefont {L.}~\bibnamefont {Chen}},\ }in\ \href {https://doi.org/10.1007/978-0-387-39940-9_133} {\emph {\bibinfo {booktitle} {{Encyclopedia of Database Systems}}}}\ (\bibinfo  {publisher} {Springer, Boston, MA},\ \bibinfo {address} {Boston, MA, USA},\ \bibinfo {year} {2009})\ pp.\ \bibinfo {pages} {545--546}\BibitemShut {NoStop}%
\bibitem [{\citenamefont {Berisha}\ \emph {et~al.}(2021)\citenamefont {Berisha}, \citenamefont {Krantsevich}, \citenamefont {Hahn}, \citenamefont {Hahn}, \citenamefont {Dasarathy}, \citenamefont {Turaga},\ and\ \citenamefont {Liss}}]{Berisha2021Oct}%
  \BibitemOpen
  \bibfield  {author} {\bibinfo {author} {\bibfnamefont {V.}~\bibnamefont {Berisha}}, \bibinfo {author} {\bibfnamefont {C.}~\bibnamefont {Krantsevich}}, \bibinfo {author} {\bibfnamefont {P.~R.}\ \bibnamefont {Hahn}}, \bibinfo {author} {\bibfnamefont {S.}~\bibnamefont {Hahn}}, \bibinfo {author} {\bibfnamefont {G.}~\bibnamefont {Dasarathy}}, \bibinfo {author} {\bibfnamefont {P.}~\bibnamefont {Turaga}},\ and\ \bibinfo {author} {\bibfnamefont {J.}~\bibnamefont {Liss}},\ }\href {https://doi.org/10.1038/s41746-021-00521-5} {\bibfield  {journal} {\bibinfo  {journal} {npj Digital Med.}\ }\textbf {\bibinfo {volume} {4}},\ \bibinfo {pages} {1} (\bibinfo {year} {2021})}\BibitemShut {NoStop}%





\bibitem [{\citenamefont {Bellman}(1961)}]{Bellman1961}%
  \BibitemOpen
  \bibfield  {author} {\bibinfo {author} {\bibfnamefont {R.~E.}\ \bibnamefont {Bellman}},\ }%
  \href@noop {} {\emph {\bibinfo {title} {Adaptive Control Processes: A Guided Tour}}}%
  \ (\bibinfo  {publisher} {Princeton University Press},\ \bibinfo {address} {Princeton, NJ},\ \bibinfo {year} {1961})%
  \BibitemShut {NoStop}%

\bibitem [{\citenamefont {Bishop}(2006)}]{Bishop2006PRML}%
  \BibitemOpen
  \bibfield  {author} {\bibinfo {author} {\bibfnamefont {C.~M.}\ \bibnamefont {Bishop}},\ }%
  \href {https://link.springer.com/book/10.1007/978-0-387-45528-0} {\emph {\bibinfo {title} {Pattern Recognition and Machine Learning}}}%
  \ (\bibinfo  {publisher} {Springer},\ \bibinfo {address} {New York, NY},\ \bibinfo {year} {2006})%
  \BibitemShut {NoStop}%

\bibitem [{\citenamefont {Verleysen}\ and\ \citenamefont {Fran{\c{c}}ois}(2005)}]{VerleysenFrancois2005Curse}%
  \BibitemOpen
  \bibfield  {author} {\bibinfo {author} {\bibfnamefont {M.}\ \bibnamefont {Verleysen}}\ and\ 
  \bibinfo {author} {\bibfnamefont {D.}\ \bibnamefont {Fran{\c{c}}ois}},\ }%
  in\ \href {https://doi.org/10.1007/11494669_93} {\emph {\bibinfo {booktitle} {Computational Intelligence and Bioinspired Systems}}}%
  \ (\bibinfo  {publisher} {Springer},\ \bibinfo {year} {2005})%
  \ pp.\ \bibinfo {pages} {758--770}%
  \BibitemShut {NoStop}%



\bibitem [{\citenamefont {Havl{\ifmmode\acute{\imath}\else\'{\i}\fi}{\ifmmode\check{c}\else\v{c}\fi}ek}\ \emph {et~al.}(2019)\citenamefont {Havl{\ifmmode\acute{\imath}\else\'{\i}\fi}{\ifmmode\check{c}\else\v{c}\fi}ek}, \citenamefont {C{\ifmmode\acute{o}\else\'{o}\fi}rcoles}, \citenamefont {Temme}, \citenamefont {Harrow}, \citenamefont {Kandala}, \citenamefont {Chow},\ and\ \citenamefont {Gambetta}}]{Havlicek2019Mar}%
  \BibitemOpen
  \bibfield  {author} {\bibinfo {author} {\bibfnamefont {V.}~\bibnamefont {Havl{\ifmmode\acute{\imath}\else\'{\i}\fi}{\ifmmode\check{c}\else\v{c}\fi}ek}}, \bibinfo {author} {\bibfnamefont {A.~D.}\ \bibnamefont {C{\ifmmode\acute{o}\else\'{o}\fi}rcoles}}, \bibinfo {author} {\bibfnamefont {K.}~\bibnamefont {Temme}}, \bibinfo {author} {\bibfnamefont {A.~W.}\ \bibnamefont {Harrow}}, \bibinfo {author} {\bibfnamefont {A.}~\bibnamefont {Kandala}}, \bibinfo {author} {\bibfnamefont {J.~M.}\ \bibnamefont {Chow}},\ and\ \bibinfo {author} {\bibfnamefont {J.~M.}\ \bibnamefont {Gambetta}},\ }\href {https://doi.org/10.1038/s41586-019-0980-2} {\bibfield  {journal} {\bibinfo  {journal} {Nature}\ }\textbf {\bibinfo {volume} {567}},\ \bibinfo {pages} {209} (\bibinfo {year} {2019})}\BibitemShut {NoStop}%
\bibitem [{\citenamefont {Peters}\ \emph {et~al.}(2021)\citenamefont {Peters}, \citenamefont {Caldeira}, \citenamefont {Ho}, \citenamefont {Leichenauer}, \citenamefont {Mohseni}, \citenamefont {Neven}, \citenamefont {Spentzouris}, \citenamefont {Strain},\ and\ \citenamefont {Perdue}}]{Peters2021Nov}%
  \BibitemOpen
  \bibfield  {author} {\bibinfo {author} {\bibfnamefont {E.}~\bibnamefont {Peters}}, \bibinfo {author} {\bibfnamefont {J.}~\bibnamefont {Caldeira}}, \bibinfo {author} {\bibfnamefont {A.}~\bibnamefont {Ho}}, \bibinfo {author} {\bibfnamefont {S.}~\bibnamefont {Leichenauer}}, \bibinfo {author} {\bibfnamefont {M.}~\bibnamefont {Mohseni}}, \bibinfo {author} {\bibfnamefont {H.}~\bibnamefont {Neven}}, \bibinfo {author} {\bibfnamefont {P.}~\bibnamefont {Spentzouris}}, \bibinfo {author} {\bibfnamefont {D.}~\bibnamefont {Strain}},\ and\ \bibinfo {author} {\bibfnamefont {G.~N.}\ \bibnamefont {Perdue}},\ }\href {https://doi.org/10.1038/s41534-021-00498-9} {\bibfield  {journal} {\bibinfo  {journal} {npj Quantum Inf.}\ }\textbf {\bibinfo {volume} {7}},\ \bibinfo {pages} {1} (\bibinfo {year} {2021})}\BibitemShut {NoStop}%
\bibitem [{\citenamefont {Kusumoto}\ \emph {et~al.}(2021)\citenamefont {Kusumoto}, \citenamefont {Mitarai}, \citenamefont {Fujii}, \citenamefont {Kitagawa},\ and\ \citenamefont {Negoro}}]{Kusumoto2021Jun}%
  \BibitemOpen
  \bibfield  {author} {\bibinfo {author} {\bibfnamefont {T.}~\bibnamefont {Kusumoto}}, \bibinfo {author} {\bibfnamefont {K.}~\bibnamefont {Mitarai}}, \bibinfo {author} {\bibfnamefont {K.}~\bibnamefont {Fujii}}, \bibinfo {author} {\bibfnamefont {M.}~\bibnamefont {Kitagawa}},\ and\ \bibinfo {author} {\bibfnamefont {M.}~\bibnamefont {Negoro}},\ }\href {https://doi.org/10.1038/s41534-021-00423-0} {\bibfield  {journal} {\bibinfo  {journal} {npj Quantum Inf.}\ }\textbf {\bibinfo {volume} {7}},\ \bibinfo {pages} {1} (\bibinfo {year} {2021})}\BibitemShut {NoStop}%
\bibitem [{\citenamefont {Wu}\ \emph {et~al.}(2021)\citenamefont {Wu}, \citenamefont {Sun}, \citenamefont {Guan}, \citenamefont {Zhou}, \citenamefont {Chan}, \citenamefont {Cheng}, \citenamefont {Pham}, \citenamefont {Qian}, \citenamefont {Wang}, \citenamefont {Zhang}, \citenamefont {Livny}, \citenamefont {Glick}, \citenamefont {Barkoutsos}, \citenamefont {Woerner}, \citenamefont {Tavernelli}, \citenamefont {Carminati}, \citenamefont {Di~Meglio}, \citenamefont {Li}, \citenamefont {Lykken}, \citenamefont {Spentzouris}, \citenamefont {Chen}, \citenamefont {Yoo},\ and\ \citenamefont {Wei}}]{Wu2021Sep}%
  \BibitemOpen
  \bibfield  {author} {\bibinfo {author} {\bibfnamefont {S.~L.}\ \bibnamefont {Wu}}, \bibinfo {author} {\bibfnamefont {S.}~\bibnamefont {Sun}}, \bibinfo {author} {\bibfnamefont {W.}~\bibnamefont {Guan}}, \bibinfo {author} {\bibfnamefont {C.}~\bibnamefont {Zhou}}, \bibinfo {author} {\bibfnamefont {J.}~\bibnamefont {Chan}}, \bibinfo {author} {\bibfnamefont {C.~L.}\ \bibnamefont {Cheng}}, \bibinfo {author} {\bibfnamefont {T.}~\bibnamefont {Pham}}, \bibinfo {author} {\bibfnamefont {Y.}~\bibnamefont {Qian}}, \bibinfo {author} {\bibfnamefont {A.~Z.}\ \bibnamefont {Wang}}, \bibinfo {author} {\bibfnamefont {R.}~\bibnamefont {Zhang}}, \bibinfo {author} {\bibfnamefont {M.}~\bibnamefont {Livny}}, \bibinfo {author} {\bibfnamefont {J.}~\bibnamefont {Glick}}, \bibinfo {author} {\bibfnamefont {P.~{\relax Kl}.}\ \bibnamefont {Barkoutsos}}, \bibinfo {author} {\bibfnamefont {S.}~\bibnamefont {Woerner}}, \bibinfo {author} {\bibfnamefont {I.}~\bibnamefont {Tavernelli}}, \bibinfo {author} {\bibfnamefont {F.}~\bibnamefont
  {Carminati}}, \bibinfo {author} {\bibfnamefont {A.}~\bibnamefont {Di~Meglio}}, \bibinfo {author} {\bibfnamefont {A.~C.~Y.}\ \bibnamefont {Li}}, \bibinfo {author} {\bibfnamefont {J.}~\bibnamefont {Lykken}}, \bibinfo {author} {\bibfnamefont {P.}~\bibnamefont {Spentzouris}}, \bibinfo {author} {\bibfnamefont {S.~Y.-C.}\ \bibnamefont {Chen}}, \bibinfo {author} {\bibfnamefont {S.}~\bibnamefont {Yoo}},\ and\ \bibinfo {author} {\bibfnamefont {T.-C.}\ \bibnamefont {Wei}},\ }\href {https://doi.org/10.1103/PhysRevResearch.3.033221} {\bibfield  {journal} {\bibinfo  {journal} {Phys. Rev. Res.}\ }\textbf {\bibinfo {volume} {3}},\ \bibinfo {pages} {033221} (\bibinfo {year} {2021})}\BibitemShut {NoStop}%
\bibitem [{\citenamefont {Alaminos}\ \emph {et~al.}(2022)\citenamefont {Alaminos}, \citenamefont {Salas},\ and\ \citenamefont {Fern{\ifmmode\acute{a}\else\'{a}\fi}ndez-G{\ifmmode\acute{a}\else\'{a}\fi}mez}}]{Alaminos2022Feb}%
  \BibitemOpen
  \bibfield  {author} {\bibinfo {author} {\bibfnamefont {D.}~\bibnamefont {Alaminos}}, \bibinfo {author} {\bibfnamefont {M.~B.}\ \bibnamefont {Salas}},\ and\ \bibinfo {author} {\bibfnamefont {M.~A.}\ \bibnamefont {Fern{\ifmmode\acute{a}\else\'{a}\fi}ndez-G{\ifmmode\acute{a}\else\'{a}\fi}mez}},\ }\href {https://doi.org/10.1007/s10614-021-10110-z} {\bibfield  {journal} {\bibinfo  {journal} {Comput. Econ.}\ }\textbf {\bibinfo {volume} {59}},\ \bibinfo {pages} {803} (\bibinfo {year} {2022})}\BibitemShut {NoStop}%
\bibitem [{\citenamefont {Lloyd}\ \emph {et~al.}(2014)\citenamefont {Lloyd}, \citenamefont {Mohseni},\ and\ \citenamefont {Rebentrost}}]{Lloyd2014Sep}%
  \BibitemOpen
  \bibfield  {author} {\bibinfo {author} {\bibfnamefont {S.}~\bibnamefont {Lloyd}}, \bibinfo {author} {\bibfnamefont {M.}~\bibnamefont {Mohseni}},\ and\ \bibinfo {author} {\bibfnamefont {P.}~\bibnamefont {Rebentrost}},\ }\href {https://doi.org/10.1038/nphys3029} {\bibfield  {journal} {\bibinfo  {journal} {Nat. Phys.}\ }\textbf {\bibinfo {volume} {10}},\ \bibinfo {pages} {631} (\bibinfo {year} {2014})}\BibitemShut {NoStop}%
\bibitem [{\citenamefont {Biamonte}\ \emph {et~al.}(2017)\citenamefont {Biamonte}, \citenamefont {Wittek}, \citenamefont {Pancotti}, \citenamefont {Rebentrost}, \citenamefont {Wiebe},\ and\ \citenamefont {Lloyd}}]{Biamonte2017Sep}%
  \BibitemOpen
  \bibfield  {author} {\bibinfo {author} {\bibfnamefont {J.}~\bibnamefont {Biamonte}}, \bibinfo {author} {\bibfnamefont {P.}~\bibnamefont {Wittek}}, \bibinfo {author} {\bibfnamefont {N.}~\bibnamefont {Pancotti}}, \bibinfo {author} {\bibfnamefont {P.}~\bibnamefont {Rebentrost}}, \bibinfo {author} {\bibfnamefont {N.}~\bibnamefont {Wiebe}},\ and\ \bibinfo {author} {\bibfnamefont {S.}~\bibnamefont {Lloyd}},\ }\href {https://doi.org/10.1038/nature23474} {\bibfield  {journal} {\bibinfo  {journal} {Nature}\ }\textbf {\bibinfo {volume} {549}},\ \bibinfo {pages} {195} (\bibinfo {year} {2017})}\BibitemShut {NoStop}%
\bibitem [{\citenamefont {Schuld}\ \emph {et~al.}(2015)\citenamefont {Schuld}, \citenamefont {Sinayskiy},\ and\ \citenamefont {Petruccione}}]{Schuld2015Apr}%
  \BibitemOpen
  \bibfield  {author} {\bibinfo {author} {\bibfnamefont {M.}~\bibnamefont {Schuld}}, \bibinfo {author} {\bibfnamefont {I.}~\bibnamefont {Sinayskiy}},\ and\ \bibinfo {author} {\bibfnamefont {F.}~\bibnamefont {Petruccione}},\ }\href {https://doi.org/10.1080/00107514.2014.964942} {\bibfield  {journal} {\bibinfo  {journal} {Contemp. Phys.}\ }\textbf {\bibinfo {volume} {56}},\ \bibinfo {pages} {172} (\bibinfo {year} {2015})}\BibitemShut {NoStop}%
\bibitem [{\citenamefont {Schuld}\ and\ \citenamefont {Killoran}(2019)}]{Schuld2019Feb}%
  \BibitemOpen
  \bibfield  {author} {\bibinfo {author} {\bibfnamefont {M.}~\bibnamefont {Schuld}}\ and\ \bibinfo {author} {\bibfnamefont {N.}~\bibnamefont {Killoran}},\ }\href {https://doi.org/10.1103/PhysRevLett.122.040504} {\bibfield  {journal} {\bibinfo  {journal} {Phys. Rev. Lett.}\ }\textbf {\bibinfo {volume} {122}},\ \bibinfo {pages} {040504} (\bibinfo {year} {2019})}\BibitemShut {NoStop}%
\bibitem [{\citenamefont {Schuld}(2021)}]{Schuld2021Jan}%
  \BibitemOpen
  \bibfield  {author} {\bibinfo {author} {\bibfnamefont {M.}~\bibnamefont {Schuld}},\ }\bibfield  {journal} {\bibinfo  {journal} {arXiv}\ }\href {https://doi.org/10.48550/arXiv.2101.11020} {10.48550/arXiv.2101.11020} (\bibinfo {year} {2021}),\ \Eprint {https://arxiv.org/abs/2101.11020} {2101.11020} \BibitemShut {NoStop}%
\bibitem [{\citenamefont {Asfaw}\ \emph {et~al.}(2020)\citenamefont {Asfaw}, \citenamefont {Bello}, \citenamefont {Ben-Haim}, \citenamefont {Bravyi}, \citenamefont {Capelluto}, \citenamefont {Vazquez}, \citenamefont {Ceroni}, \citenamefont {Chen}, \citenamefont {Frisch}, \citenamefont {Gambetta}, \citenamefont {Garion}, \citenamefont {Gil}, \citenamefont {Gonzalez}, \citenamefont {Harkins}, \citenamefont {Imamichi}, \citenamefont {McKay}, \citenamefont {Mezzacapo}, \citenamefont {Minev}, \citenamefont {Movassagh}, \citenamefont {Nannicini}, \citenamefont {Nation}, \citenamefont {Phan}, \citenamefont {Pistoia}, \citenamefont {Rattew}, \citenamefont {Schaefer}, \citenamefont {Shabani}, \citenamefont {Smolin}, \citenamefont {Temme}, \citenamefont {Tod},\ and\ \citenamefont {Wood}}]{qiskit-textbook}%
  \BibitemOpen
  \bibfield  {author} {\bibinfo {author} {\bibfnamefont {A.}~\bibnamefont {Asfaw}}, \bibinfo {author} {\bibfnamefont {L.}~\bibnamefont {Bello}}, \bibinfo {author} {\bibfnamefont {Y.}~\bibnamefont {Ben-Haim}}, \bibinfo {author} {\bibfnamefont {S.}~\bibnamefont {Bravyi}}, \bibinfo {author} {\bibfnamefont {L.}~\bibnamefont {Capelluto}}, \bibinfo {author} {\bibfnamefont {A.~C.}\ \bibnamefont {Vazquez}}, \bibinfo {author} {\bibfnamefont {J.}~\bibnamefont {Ceroni}}, \bibinfo {author} {\bibfnamefont {R.}~\bibnamefont {Chen}}, \bibinfo {author} {\bibfnamefont {A.}~\bibnamefont {Frisch}}, \bibinfo {author} {\bibfnamefont {J.}~\bibnamefont {Gambetta}}, \bibinfo {author} {\bibfnamefont {S.}~\bibnamefont {Garion}}, \bibinfo {author} {\bibfnamefont {L.}~\bibnamefont {Gil}}, \bibinfo {author} {\bibfnamefont {S.~D. L.~P.}\ \bibnamefont {Gonzalez}}, \bibinfo {author} {\bibfnamefont {F.}~\bibnamefont {Harkins}}, \bibinfo {author} {\bibfnamefont {T.}~\bibnamefont {Imamichi}}, \bibinfo {author} {\bibfnamefont {D.}~\bibnamefont
  {McKay}}, \bibinfo {author} {\bibfnamefont {A.}~\bibnamefont {Mezzacapo}}, \bibinfo {author} {\bibfnamefont {Z.}~\bibnamefont {Minev}}, \bibinfo {author} {\bibfnamefont {R.}~\bibnamefont {Movassagh}}, \bibinfo {author} {\bibfnamefont {G.}~\bibnamefont {Nannicini}}, \bibinfo {author} {\bibfnamefont {P.}~\bibnamefont {Nation}}, \bibinfo {author} {\bibfnamefont {A.}~\bibnamefont {Phan}}, \bibinfo {author} {\bibfnamefont {M.}~\bibnamefont {Pistoia}}, \bibinfo {author} {\bibfnamefont {A.}~\bibnamefont {Rattew}}, \bibinfo {author} {\bibfnamefont {J.}~\bibnamefont {Schaefer}}, \bibinfo {author} {\bibfnamefont {J.}~\bibnamefont {Shabani}}, \bibinfo {author} {\bibfnamefont {J.}~\bibnamefont {Smolin}}, \bibinfo {author} {\bibfnamefont {K.}~\bibnamefont {Temme}}, \bibinfo {author} {\bibfnamefont {M.}~\bibnamefont {Tod}},\ and\ \bibinfo {author} {\bibfnamefont {S.}~\bibnamefont {Wood}},\ }\href {https://qiskit.org/textbook/preface.html} {\emph {\bibinfo {title} {Learn Quantum Computation Using Qiskit}}}\ (\bibinfo
  {year} {2020})\BibitemShut {NoStop}%
\bibitem [{\citenamefont {Cong}\ \emph {et~al.}(2019)\citenamefont {Cong}, \citenamefont {Choi},\ and\ \citenamefont {Lukin}}]{cong2019quantum}%
  \BibitemOpen
  \bibfield  {author} {\bibinfo {author} {\bibfnamefont {I.}~\bibnamefont {Cong}}, \bibinfo {author} {\bibfnamefont {S.}~\bibnamefont {Choi}},\ and\ \bibinfo {author} {\bibfnamefont {M.~D.}\ \bibnamefont {Lukin}},\ }\href {https://doi.org/10.1038/s41567-019-0648-8} {\bibfield  {journal} {\bibinfo  {journal} {Nat. Phys.}\ }\textbf {\bibinfo {volume} {15}},\ \bibinfo {pages} {1273} (\bibinfo {year} {2019})}\BibitemShut {NoStop}%
\bibitem [{\citenamefont {Zoufal}\ \emph {et~al.}(2019)\citenamefont {Zoufal}, \citenamefont {Lucchi},\ and\ \citenamefont {Woerner}}]{zoufal2019quantum}%
  \BibitemOpen
  \bibfield  {author} {\bibinfo {author} {\bibfnamefont {C.}~\bibnamefont {Zoufal}}, \bibinfo {author} {\bibfnamefont {A.}~\bibnamefont {Lucchi}},\ and\ \bibinfo {author} {\bibfnamefont {S.}~\bibnamefont {Woerner}},\ }\href {https://doi.org/10.1038/s41534-019-0223-2} {\bibfield  {journal} {\bibinfo  {journal} {npj Quantum Inf.}\ }\textbf {\bibinfo {volume} {5}},\ \bibinfo {pages} {1} (\bibinfo {year} {2019})}\BibitemShut {NoStop}%
\bibitem [{\citenamefont {P{\'{e}}rez-Salinas}\ \emph {et~al.}(2020)\citenamefont {P{\'{e}}rez-Salinas}, \citenamefont {Cervera-Lierta}, \citenamefont {Gil-Fuster},\ and\ \citenamefont {Latorre}}]{PerezSalinas2020datareuploading}%
  \BibitemOpen
  \bibfield  {author} {\bibinfo {author} {\bibfnamefont {A.}~\bibnamefont {P{\'{e}}rez-Salinas}}, \bibinfo {author} {\bibfnamefont {A.}~\bibnamefont {Cervera-Lierta}}, \bibinfo {author} {\bibfnamefont {E.}~\bibnamefont {Gil-Fuster}},\ and\ \bibinfo {author} {\bibfnamefont {J.~I.}\ \bibnamefont {Latorre}},\ }\href {https://doi.org/10.22331/q-2020-02-06-226} {\bibfield  {journal} {\bibinfo  {journal} {{Quantum}}\ }\textbf {\bibinfo {volume} {4}},\ \bibinfo {pages} {226} (\bibinfo {year} {2020})}\BibitemShut {NoStop}%
\bibitem [{\citenamefont {Roy}()}]{encoding}%
  \BibitemOpen
  \bibfield  {author} {\bibinfo {author} {\bibfnamefont {B.}~\bibnamefont {Roy}},\ }\href {https://medium.datadriveninvestor.com/all-about-data-encoding-for-quantum-machine-learning-2a7344b1dfef} {\bibinfo {title} {All about data encoding for quantum machine learning}}\BibitemShut {NoStop}%
\bibitem [{\citenamefont {Preskill}(2018)}]{Preskill2018Aug}%
  \BibitemOpen
  \bibfield  {author} {\bibinfo {author} {\bibfnamefont {J.}~\bibnamefont {Preskill}},\ }\href {https://doi.org/10.22331/q-2018-08-06-79} {\bibfield  {journal} {\bibinfo  {journal} {Quantum}\ }\textbf {\bibinfo {volume} {2}},\ \bibinfo {pages} {79} (\bibinfo {year} {2018})},\ \Eprint {https://arxiv.org/abs/1801.00862v3} {1801.00862v3} \BibitemShut {NoStop}%
\bibitem [{\citenamefont {Hubregtsen}\ \emph {et~al.}(2021)\citenamefont {Hubregtsen}, \citenamefont {Pichlmeier}, \citenamefont {Stecher},\ and\ \citenamefont {Bertels}}]{Hubregtsen2021Mar}%
  \BibitemOpen
  \bibfield  {author} {\bibinfo {author} {\bibfnamefont {T.}~\bibnamefont {Hubregtsen}}, \bibinfo {author} {\bibfnamefont {J.}~\bibnamefont {Pichlmeier}}, \bibinfo {author} {\bibfnamefont {P.}~\bibnamefont {Stecher}},\ and\ \bibinfo {author} {\bibfnamefont {K.}~\bibnamefont {Bertels}},\ }\href {https://doi.org/10.1007/s42484-021-00038-w} {\bibfield  {journal} {\bibinfo  {journal} {Quantum Mach. Intell.}\ }\textbf {\bibinfo {volume} {3}},\ \bibinfo {pages} {9} (\bibinfo {year} {2021})}\BibitemShut {NoStop}%
\bibitem [{\citenamefont {{Qiskit contributors}}(2023)}]{Qiskit}%
  \BibitemOpen
  \bibfield  {author} {\bibinfo {author} {\bibnamefont {{Qiskit contributors}}},\ }\href {https://doi.org/10.5281/zenodo.2573505} {\bibinfo {title} {Qiskit: An open-source framework for quantum computing}} (\bibinfo {year} {2023})\BibitemShut {NoStop}%
\bibitem [{\citenamefont {Vasques}\ \emph {et~al.}(2023)\citenamefont {Vasques}, \citenamefont {Paik},\ and\ \citenamefont {Cif}}]{Vasques2023Jul}%
  \BibitemOpen
  \bibfield  {author} {\bibinfo {author} {\bibfnamefont {X.}~\bibnamefont {Vasques}}, \bibinfo {author} {\bibfnamefont {H.}~\bibnamefont {Paik}},\ and\ \bibinfo {author} {\bibfnamefont {L.}~\bibnamefont {Cif}},\ }\href {https://doi.org/10.1038/s41598-023-38558-z} {\bibfield  {journal} {\bibinfo  {journal} {Sci. Rep.}\ }\textbf {\bibinfo {volume} {13}},\ \bibinfo {pages} {1} (\bibinfo {year} {2023})}\BibitemShut {NoStop}%
\bibitem [{\citenamefont {Kim}\ \emph {et~al.}()\citenamefont {Kim}, \citenamefont {Song}, \citenamefont {Jang},\ and\ \citenamefont {Seo}}]{ZZKim}%
  \BibitemOpen
  \bibfield  {author} {\bibinfo {author} {\bibfnamefont {H.-J.}\ \bibnamefont {Kim}}, \bibinfo {author} {\bibfnamefont {G.-J.}\ \bibnamefont {Song}}, \bibinfo {author} {\bibfnamefont {K.-B.}\ \bibnamefont {Jang}},\ and\ \bibinfo {author} {\bibfnamefont {H.-J.}\ \bibnamefont {Seo}},\ }in\ \href {https://doi.org/10.1109/ICCE-Asia53811.2021.9641932} {\emph {\bibinfo {booktitle} {{2021 IEEE International Conference on Consumer Electronics-Asia (ICCE-Asia)}}}}\ (\bibinfo  {publisher} {IEEE})\ pp.\ \bibinfo {pages} {01--03}\BibitemShut {NoStop}%
\bibitem [{\citenamefont {Abbas}\ \emph {et~al.}(2021)\citenamefont {Abbas}, \citenamefont {Sutter}, \citenamefont {Zoufal}, \citenamefont {Lucchi}, \citenamefont {Figalli},\ and\ \citenamefont {Woerner}}]{ZZAbbas2021Jun}%
  \BibitemOpen
  \bibfield  {author} {\bibinfo {author} {\bibfnamefont {A.}~\bibnamefont {Abbas}}, \bibinfo {author} {\bibfnamefont {D.}~\bibnamefont {Sutter}}, \bibinfo {author} {\bibfnamefont {C.}~\bibnamefont {Zoufal}}, \bibinfo {author} {\bibfnamefont {A.}~\bibnamefont {Lucchi}}, \bibinfo {author} {\bibfnamefont {A.}~\bibnamefont {Figalli}},\ and\ \bibinfo {author} {\bibfnamefont {S.}~\bibnamefont {Woerner}},\ }\href {https://doi.org/10.1038/s43588-021-00084-1} {\bibfield  {journal} {\bibinfo  {journal} {Nat. Comput. Sci.}\ }\textbf {\bibinfo {volume} {1}},\ \bibinfo {pages} {403} (\bibinfo {year} {2021})}\BibitemShut {NoStop}%
\bibitem [{\citenamefont {Mukhamedieva}(2024)}]{ZZMukhamedieva2024}%
  \BibitemOpen
  \bibfield  {author} {\bibinfo {author} {\bibfnamefont {D.~T.}\ \bibnamefont {Mukhamedieva}},\ }\href {https://doi.org/10.1051/e3sconf/202449404026} {\bibfield  {journal} {\bibinfo  {journal} {E3S Web Conf.}\ }\textbf {\bibinfo {volume} {494}},\ \bibinfo {pages} {04026} (\bibinfo {year} {2024})}\BibitemShut {NoStop}%
\bibitem [{Bib(2024{\natexlab{a}})}]{BibEntry2024Jan}%
  \BibitemOpen
  \href {https://www.combinatorics.org/ojs/index.php/eljc/article/view/v13i1r26/pdf} {\bibinfo {title} {{View of Meta-Fibonacci Sequences, Binary Trees and Extremal Compact Codes}}} (\bibinfo {year} {2024}{\natexlab{a}}),\ \bibinfo {note} {[Online; accessed 22. Jan. 2024]}\BibitemShut {NoStop}%
\bibitem [{Bib(2024{\natexlab{b}})}]{BibEntry2024Jan2}%
  \BibitemOpen
  \href {https://oeis.org/search?q=1%2C2%2C2%2C3%2C+4%2C+4%2C+4%2C+5%2C+6%2C+6%2C+7%2C+8%2C+8%2C+8%2C+8%2C+9%2C+10%2C+10%2C+11%2C+12%2C+12%2C+12%2C+13%2C+14%2C+14%2C+15%2C+16%2C+16%2C+16%2C+16%2C+16%2C+17&go=Search} {\bibinfo {title} {{1,2,2,3, 4, 4, 4, 5, 6, 6, 7, 8, 8, 8, 8, 9, 10, 10, 11, 12, 12, 12, 13, 14, 14, 15, 16, 16, 16, 16, 16, 17 - OEIS}}} (\bibinfo {year} {2024}{\natexlab{b}}),\ \bibinfo {note} {[Online; accessed 22. Jan. 2024]}\BibitemShut {NoStop}%
\bibitem [{\citenamefont {Vatan}\ and\ \citenamefont {Williams}(2004)}]{Vatan2004Mar}%
  \BibitemOpen
  \bibfield  {author} {\bibinfo {author} {\bibfnamefont {F.}~\bibnamefont {Vatan}}\ and\ \bibinfo {author} {\bibfnamefont {C.}~\bibnamefont {Williams}},\ }\href {https://doi.org/10.1103/PhysRevA.69.032315} {\bibfield  {journal} {\bibinfo  {journal} {Phys. Rev. A}\ }\textbf {\bibinfo {volume} {69}},\ \bibinfo {pages} {032315} (\bibinfo {year} {2004})}\BibitemShut {NoStop}%
\bibitem [{\citenamefont {Pedregosa}\ \emph {et~al.}(2011)\citenamefont {Pedregosa}, \citenamefont {Varoquaux}, \citenamefont {Gramfort}, \citenamefont {Michel}, \citenamefont {Thirion}, \citenamefont {Grisel}, \citenamefont {Blondel}, \citenamefont {Prettenhofer}, \citenamefont {Weiss}, \citenamefont {Dubourg}, \citenamefont {Vanderplas}, \citenamefont {Passos}, \citenamefont {Cournapeau}, \citenamefont {Brucher}, \citenamefont {Perrot},\ and\ \citenamefont {Duchesnay}}]{scikit-learn}%
  \BibitemOpen
  \bibfield  {author} {\bibinfo {author} {\bibfnamefont {F.}~\bibnamefont {Pedregosa}}, \bibinfo {author} {\bibfnamefont {G.}~\bibnamefont {Varoquaux}}, \bibinfo {author} {\bibfnamefont {A.}~\bibnamefont {Gramfort}}, \bibinfo {author} {\bibfnamefont {V.}~\bibnamefont {Michel}}, \bibinfo {author} {\bibfnamefont {B.}~\bibnamefont {Thirion}}, \bibinfo {author} {\bibfnamefont {O.}~\bibnamefont {Grisel}}, \bibinfo {author} {\bibfnamefont {M.}~\bibnamefont {Blondel}}, \bibinfo {author} {\bibfnamefont {P.}~\bibnamefont {Prettenhofer}}, \bibinfo {author} {\bibfnamefont {R.}~\bibnamefont {Weiss}}, \bibinfo {author} {\bibfnamefont {V.}~\bibnamefont {Dubourg}}, \bibinfo {author} {\bibfnamefont {J.}~\bibnamefont {Vanderplas}}, \bibinfo {author} {\bibfnamefont {A.}~\bibnamefont {Passos}}, \bibinfo {author} {\bibfnamefont {D.}~\bibnamefont {Cournapeau}}, \bibinfo {author} {\bibfnamefont {M.}~\bibnamefont {Brucher}}, \bibinfo {author} {\bibfnamefont {M.}~\bibnamefont {Perrot}},\ and\ \bibinfo {author} {\bibfnamefont
  {E.}~\bibnamefont {Duchesnay}},\ }\href@noop {} {\bibfield  {journal} {\bibinfo  {journal} {Journal of Machine Learning Research}\ }\textbf {\bibinfo {volume} {12}},\ \bibinfo {pages} {2825} (\bibinfo {year} {2011})}\BibitemShut {NoStop}%
\bibitem [{\citenamefont {Chicco}(2017)}]{Chicco2017}%
  \BibitemOpen
  \bibfield  {author} {\bibinfo {author} {\bibfnamefont {D.}~\bibnamefont {Chicco}},\ }\href {https://doi.org/10.1186/s13040-017-0155-3} {\bibfield  {journal} {\bibinfo  {journal} {BioData Mining}\ }\textbf {\bibinfo {volume} {10}},\ \bibinfo {pages} {1} (\bibinfo {year} {2017})}\BibitemShut {NoStop}%
\bibitem [{\citenamefont {Fawcett}(2006)}]{Fawcett2006}%
  \BibitemOpen
  \bibfield  {author} {\bibinfo {author} {\bibfnamefont {T.}~\bibnamefont {Fawcett}},\ }\href {https://doi.org/10.1016/j.patrec.2005.10.010} {\bibfield  {journal} {\bibinfo  {journal} {Pattern Recognit. Lett.}\ }\textbf {\bibinfo {volume} {27}},\ \bibinfo {pages} {861} (\bibinfo {year} {2006})}\BibitemShut {NoStop}%
\bibitem [{\citenamefont {Chicco}\ and\ \citenamefont {Jurman}(2020)}]{Chicco2020Dec}%
  \BibitemOpen
  \bibfield  {author} {\bibinfo {author} {\bibfnamefont {D.}~\bibnamefont {Chicco}}\ and\ \bibinfo {author} {\bibfnamefont {G.}~\bibnamefont {Jurman}},\ }\href {https://doi.org/10.1186/s12864-019-6413-7} {\bibfield  {journal} {\bibinfo  {journal} {BMC Genomics}\ }\textbf {\bibinfo {volume} {21}},\ \bibinfo {pages} {1} (\bibinfo {year} {2020})}\BibitemShut {NoStop}%
\bibitem [{\citenamefont {Japkowicz}\ and\ \citenamefont {Stephen}(2002)}]{japkowicz2002}%
  \BibitemOpen
  \bibfield  {author} {\bibinfo {author} {\bibfnamefont {N.}~\bibnamefont {Japkowicz}}\ and\ \bibinfo {author} {\bibfnamefont {S.}~\bibnamefont {Stephen}},\ }\href {https://doi.org/10.3233/IDA-2002-6504} {\bibfield  {journal} {\bibinfo  {journal} {Intell. Data Anal.}\ }\textbf {\bibinfo {volume} {6}},\ \bibinfo {pages} {429} (\bibinfo {year} {2002})}\BibitemShut {NoStop}%
\bibitem [{\citenamefont {Jeni}\ \emph {et~al.}(2013)\citenamefont {Jeni}, \citenamefont {Cohn},\ and\ \citenamefont {De~La~Torre}}]{Jeni2013}%
  \BibitemOpen
  \bibfield  {author} {\bibinfo {author} {\bibfnamefont {L.~A.}\ \bibnamefont {Jeni}}, \bibinfo {author} {\bibfnamefont {J.~F.}\ \bibnamefont {Cohn}},\ and\ \bibinfo {author} {\bibfnamefont {F.}~\bibnamefont {De~La~Torre}},\ }in\ \href {https://doi.org/10.1109/ACII.2013.47} {\emph {\bibinfo {booktitle} {{2013 Humaine Association Conference on Affective Computing and Intelligent Interaction}}}}\ (\bibinfo  {publisher} {IEEE},\ \bibinfo {year} {2013})\ pp.\ \bibinfo {pages} {245--251}\BibitemShut {NoStop}%
\bibitem [{\citenamefont {Sokolova}\ and\ \citenamefont {Lapalme}(2009)}]{sokolova2009}%
  \BibitemOpen
  \bibfield  {author} {\bibinfo {author} {\bibfnamefont {M.}~\bibnamefont {Sokolova}}\ and\ \bibinfo {author} {\bibfnamefont {G.}~\bibnamefont {Lapalme}},\ }\href {https://doi.org/10.1016/j.ipm.2009.03.002} {\bibfield  {journal} {\bibinfo  {journal} {Information Processing {\&} Management}\ }\textbf {\bibinfo {volume} {45}},\ \bibinfo {pages} {427} (\bibinfo {year} {2009})}\BibitemShut {NoStop}%

\bibitem [{\citenamefont {Thanasilp}\ \citenamefont {Wang}\ \citenamefont {Cerezo}\ and\ \citenamefont {Holmes}(2024)}]{thanasilp2024}%
  \BibitemOpen
  \bibfield  {author} {\bibinfo {author} {\bibfnamefont {S.}~\bibnamefont {Thanasilp}}, 
  \bibinfo {author} {\bibfnamefont {S.}~\bibnamefont {Wang}}, 
  \bibinfo {author} {\bibfnamefont {M.}~\bibnamefont {Cerezo}},\ and\ 
  \bibinfo {author} {\bibfnamefont {Z.}~\bibnamefont {Holmes}},\ }\href {https://doi.org/10.1038/s41467-024-49287-w} 
  {\bibfield  {journal} {\bibinfo  {journal} {Nature Communications}\ }\textbf {\bibinfo {volume} {15}},\ \bibinfo {pages} {5200} (\bibinfo {year} {2024})}\BibitemShut {NoStop}%



\bibitem [{\citenamefont {Jha}\ \emph {et~al.}(2026)\citenamefont {Jha}, \citenamefont {Kasabov}, \citenamefont {Bhattacharyya}, \citenamefont {Coyle},\ and\ \citenamefont {Prasad}}]{nature2026comparative}%
\BibitemOpen
\bibfield  {author} {\bibinfo {author} {\bibfnamefont {R.~K.}\ \bibnamefont {Jha}}, \bibinfo {author} {\bibfnamefont {N.}\ \bibnamefont {Kasabov}}, \bibinfo {author} {\bibfnamefont {S.}\ \bibnamefont {Bhattacharyya}}, \bibinfo {author} {\bibfnamefont {D.}\ \bibnamefont {Coyle}},\ and\ \bibinfo {author} {\bibfnamefont {G.}\ \bibnamefont {Prasad}},\ }\href {https://doi.org/10.1038/s41598-026-39392-9} {\bibfield  {journal} {\bibinfo  {journal} {Sci. Rep.}\ }\textbf {\bibinfo {volume} {16}},\ \bibinfo {pages} {8142} (\bibinfo {year} {2026})}\BibitemShut {NoStop}%

\bibitem [{\citenamefont {El~Hafidi}\ \emph {et~al.}(2025)\citenamefont {El~Hafidi}, \citenamefont {Toufah},\ and\ \citenamefont {Kadim}}]{lungcancer2025qsvm}%
\BibitemOpen
\bibfield  {author} {\bibinfo {author} {\bibfnamefont {M.~Y.}\ \bibnamefont {El~Hafidi}}, \bibinfo {author} {\bibfnamefont {A.}\ \bibnamefont {Toufah}},\ and\ \bibinfo {author} {\bibfnamefont {M.~A.}\ \bibnamefont {Kadim}},\ }\href {https://dx.doi.org/10.29328/journal.jairi.1001012} {\bibinfo  {note} {Online resource, 2025}}\BibitemShut {NoStop}%

\bibitem [{\citenamefont {Akpinar}(2024)}]{alzheimer2023qml}%
\BibitemOpen
\bibfield  {author} {\bibinfo {author} {\bibfnamefont {E.}\ \bibnamefont {Akpinar}},\ }\href {https://doi.org/10.1109/HPEC62836.2024.10938482} {\bibfield  {journal} {\bibinfo  {journal} {Proc. IEEE HPEC}\ }\ \bibinfo {pages} {1--6} (\bibinfo {year} {2024})}\BibitemShut {NoStop}%

\bibitem [{\citenamefont {Alvarez-Estevez}(2025)}]{benchmarking2024kernel}%
\BibitemOpen
\bibfield  {author} {\bibinfo {author} {\bibfnamefont {D.}\ \bibnamefont {Alvarez-Estevez}},\ }\href {https://doi.org/10.1109/TQE.2025.3541882} {\bibfield  {journal} {\bibinfo  {journal} {IEEE Trans. Quantum Eng.}\ }\textbf {\bibinfo {volume} {6}},\ \bibinfo {pages} {1--15} (\bibinfo {year} {2025})}\BibitemShut {NoStop}%

\bibitem [{\citenamefont {Boy}\ \emph {et~al.}(2025)\citenamefont {Boy}, \citenamefont {Altamura}, \citenamefont {Manawadu}, \citenamefont {Tavernelli}, \citenamefont {Mensa},\ and\ \citenamefont {Wales}}]{Boy_2025}%
\BibitemOpen
\bibfield  {author} {\bibinfo {author} {\bibfnamefont {C.}\ \bibnamefont {Boy}}, \bibinfo {author} {\bibfnamefont {E.}\ \bibnamefont {Altamura}}, \bibinfo {author} {\bibfnamefont {D.}\ \bibnamefont {Manawadu}}, \bibinfo {author} {\bibfnamefont {I.}\ \bibnamefont {Tavernelli}}, \bibinfo {author} {\bibfnamefont {S.}\ \bibnamefont {Mensa}},\ and\ \bibinfo {author} {\bibfnamefont {D.~J.}\ \bibnamefont {Wales}},\ }\href {https://doi.org/10.1088/2632-2153/ae304f} {\bibfield  {journal} {\bibinfo  {journal} {Mach. Learn.: Sci. Technol.}\ }\textbf {\bibinfo {volume} {6}},\ \bibinfo {pages} {045076} (\bibinfo {year} {2025})}\BibitemShut {NoStop}%

\bibitem [{\citenamefont {Chandrasekhar}\ \emph {et~al.}(2025)\citenamefont {Chandrasekhar}, \citenamefont {Pokhrel},\ and\ \citenamefont {Singh}}]{bonddissociation2025adapting}%
\BibitemOpen
\bibfield  {author} {\bibinfo {author} {\bibfnamefont {S.}\ \bibnamefont {Chandrasekhar}}, \bibinfo {author} {\bibfnamefont {S.~R.}\ \bibnamefont {Pokhrel}},\ and\ \bibinfo {author} {\bibfnamefont {N.}\ \bibnamefont {Singh}},\ }\href {https://arxiv.org/abs/2510.06563} {\bibinfo  {journal} {arXiv:2510.06563} (\bibinfo {year} {2025})}\BibitemShut {NoStop}%

\bibitem [{\citenamefont {Novák}\ \emph {et~al.}(2025)\citenamefont {Novák}, \citenamefont {Zelinka}, \citenamefont {Přibylová}, \citenamefont {Martínek},\ and\ \citenamefont {Benčurik}}]{anastomoticleak2025qml}%
\BibitemOpen
\bibfield  {author} {\bibinfo {author} {\bibfnamefont {V.}\ \bibnamefont {Novák}}, \bibinfo {author} {\bibfnamefont {I.}\ \bibnamefont {Zelinka}}, \bibinfo {author} {\bibfnamefont {L.}\ \bibnamefont {Přibylová}}, \bibinfo {author} {\bibfnamefont {L.}\ \bibnamefont {Martínek}},\ and\ \bibinfo {author} {\bibfnamefont {V.}\ \bibnamefont {Benčurik}},\ }\href {https://arxiv.org/abs/2506.01708} {\bibinfo  {journal} {arXiv:2506.01708} (\bibinfo {year} {2025})}\BibitemShut {NoStop}%

\bibitem [{\citenamefont {Choi}\ \emph {et~al.}(2024)\citenamefont {Choi}, \citenamefont {Sul}, \citenamefont {Kim}, \citenamefont {Hong}, \citenamefont {Izquierdo~Gonzalez}, \citenamefont {Cembellin},\ and\ \citenamefont {Wang}}]{anomaly2024additive}%
\BibitemOpen
\bibfield  {author} {\bibinfo {author} {\bibfnamefont {E.}\ \bibnamefont {Choi}}, \bibinfo {author} {\bibfnamefont {J.}\ \bibnamefont {Sul}}, \bibinfo {author} {\bibfnamefont {J.~E.}\ \bibnamefont {Kim}}, \bibinfo {author} {\bibfnamefont {S.}\ \bibnamefont {Hong}}, \bibinfo {author} {\bibfnamefont {B.}\ \bibnamefont {Izquierdo~Gonzalez}}, \bibinfo {author} {\bibfnamefont {P.}\ \bibnamefont {Cembellin}},\ and\ \bibinfo {author} {\bibfnamefont {Y.}\ \bibnamefont {Wang}},\ }\href {https://doi.org/10.1016/j.mfglet.2024.09.168} {\bibfield  {journal} {\bibinfo  {journal} {Manuf. Lett.}\ }\textbf {\bibinfo {volume} {41}},\ \bibinfo {pages} {1415--1422} (\bibinfo {year} {2024})}\BibitemShut {NoStop}%

\bibitem [{\citenamefont {Kwon}\ \emph {et~al.}(2025)\citenamefont {Kwon}, \citenamefont {Huh}, \citenamefont {Kwon}, \citenamefont {Choi},\ and\ \citenamefont {Kwon}}]{imbalance2025leveraging}%
\BibitemOpen
\bibfield  {author} {\bibinfo {author} {\bibfnamefont {S.}\ \bibnamefont {Kwon}}, \bibinfo {author} {\bibfnamefont {J.}\ \bibnamefont {Huh}}, \bibinfo {author} {\bibfnamefont {S.~J.}\ \bibnamefont {Kwon}}, \bibinfo {author} {\bibfnamefont {S.-H.}\ \bibnamefont {Choi}},\ and\ \bibinfo {author} {\bibfnamefont {O.}\ \bibnamefont {Kwon}},\ }\href {https://doi.org/10.3390/sym17020186} {\bibfield  {journal} {\bibinfo  {journal} {Symmetry}\ }\textbf {\bibinfo {volume} {17}},\ \bibinfo {pages} {186} (\bibinfo {year} {2025})}\BibitemShut {NoStop}%

\bibitem [{\citenamefont {Innan}\ \emph {et~al.}(2024)\citenamefont {Innan}, \citenamefont {Khan},\ and\ \citenamefont {Bennai}}]{comparative2023study}%
\BibitemOpen
\bibfield  {author} {\bibinfo {author} {\bibfnamefont {N.}\ \bibnamefont {Innan}}, \bibinfo {author} {\bibfnamefont {M.~A.-Z.}\ \bibnamefont {Khan}},\ and\ \bibinfo {author} {\bibfnamefont {M.}\ \bibnamefont {Bennai}},\ }\href {https://doi.org/10.1142/S0219749923500442} {\bibfield  {journal} {\bibinfo  {journal} {Int. J. Quantum Inf.}\ }\textbf {\bibinfo {volume} {22}},\ \bibinfo {pages} {2350044} (\bibinfo {year} {2024})}\BibitemShut {NoStop}%

\bibitem [{\citenamefont {Singh}\ and\ \citenamefont {Pokhrel}(2025)}]{genomic2025modeling}%
\BibitemOpen
\bibfield  {author} {\bibinfo {author} {\bibfnamefont {N.}\ \bibnamefont {Singh}}\ and\ \bibinfo {author} {\bibfnamefont {S.~R.}\ \bibnamefont {Pokhrel}},\ }\href {https://doi.org/10.1109/TAI.2025.3630124} {\bibfield  {journal} {\bibinfo  {journal} {IEEE Trans. Artif. Intell.}\ }\ \bibinfo {pages} {1--13} (\bibinfo {year} {2025})}\BibitemShut {NoStop}%


\bibitem [{\citenamefont {Bermejo}\ \emph {et~al.}(2026)\citenamefont {Bermejo}, \citenamefont {Braccia}, \citenamefont {Rudolph}, \citenamefont {Holmes}, \citenamefont {Cincio},\ and\ \citenamefont {Cerezo}}]{bermejo_simulable}%
\BibitemOpen
\bibfield  {author} {\bibinfo {author} {\bibfnamefont {P.}\ \bibnamefont {Bermejo}}, \bibinfo {author} {\bibfnamefont {P.}\ \bibnamefont {Braccia}}, \bibinfo {author} {\bibfnamefont {M.~S.}\ \bibnamefont {Rudolph}}, \bibinfo {author} {\bibfnamefont {Z.}\ \bibnamefont {Holmes}}, \bibinfo {author} {\bibfnamefont {L.}\ \bibnamefont {Cincio}},\ and\ \bibinfo {author} {\bibfnamefont {M.}\ \bibnamefont {Cerezo}},\ }\href {https://link.aps.org/doi/10.1103/8qt9-72ts} {\bibfield  {journal} {\bibinfo  {journal} {PRX Quantum}\ }\ (\bibinfo {year} {2026})}\BibitemShut {NoStop}%

\end{thebibliography}
\end{document}